\input amstex
\magnification=1200
\documentstyle{amsppt}
\TagsOnRight
\NoBlackBoxes

\def\pd#1#2{\frac{\partial{#1}}{\partial{#2}}} 
\def\ss{\vskip .3cm}
\def\ms{\vskip .5cm}
\def\ni{\noindent}
\def\tt{\rightarrow}

\def\inv{^{-1}}

\def\tti{\tt \infty}
\def\tnin{\tan^{-1}}

\def\D{\Delta}
\def\t{\Theta}
\def\e{\epsilon}
\def\k{\kappa}
\def\z{\zeta}
\def\l{\lambda}
\def\g{\gamma}
\def\r{\rho}
\def\s{\sigma}
\def\Psid{\Psi^{\dagger}}

\def\rs{\rho^*_{0,t}}
\def\te{\tilde\eta}
\def\intr{\int^{\infty}_{-\infty}}
\def\la{\langle}
\def\ra{\rangle}

\def\R{\mathop{{\text{I}}\mskip-4.0mu{\text{R}}}\nolimits}

\topmatter
\centerline{\bf A DIRAC SEA AND THERMODYNAMIC EQUILIBRIUM}

\centerline{\bf FOR THE QUANTIZED THREE WAVE INTERACTION}

\ms
Richard Beals 

\sl Yale University, Department of Mathematics, Box 208283

New Haven, Connecticut 06520-8283
\smallskip
\rm David Sattinger

\sl University of Minnesota, Department of Mathematics

Minneapolis, MN  55455
\smallskip
\rm Eric Williams

\sl University of Tokyo, Institute for Solid State Physics

7-22-1 Roppongi, Tokyo 1106 \rm

\abstract{The classical version of the three wave interaction models 
the creation and destruction of waves; the quantized
version models the creation and destruction of particles. The quantum
three wave interaction is described and the Bethe Ansatz for the
eigenfunctions is given in closed form.  The Bethe equations are
derived in a rigorous fashion and are shown to have a thermodynamic
limit.  The Dirac sea of negative energy states is obtained as  
the infinite density limit.  Finite particle/hole excitations are
determined and the asymptotic relation of energy and momentum is
obtained.  The Yang-Yang functional for the relative
free energy of finite density excitations is constructed
and is shown to be convex and bounded below.  The equations of
thermal equilibrium are obtained.}
\endabstract

\endtopmatter

\ni
{\bf 1. INTRODUCTION}
\ss
The creation and destruction of particles is a fundamental process in
high energy physics.  Among other things, the presence of such
processes implies that the entries of the scattering matrix between
states with different numbers of elementary excitations (or physical 
particles) can be non-zero.  

Integrable systems provide solvable paradigms for various physical
phenomena: {\it e.g.} for lattice statistical models, the algebraic decay of
correlation functions at a critical point may be observed.  The creation
and destruction of particle pairs is modeled by the quantum
three wave interaction (3WI), which is obtained by
quantizing a particular reduction of the $N=3$ case of the AKNS
hierarchy as in Wadati and Ohkuma$^1$,
 or by taking the continuum limit of the $GL(3)$
lattice magnet as in Kulish and Reshetikhin$^2$.  The result is an 
interacting many body system which does not conserve the number of 
particles: the interaction permits two distinct types of particles 
to combine to form a third, and also the reverse process.

The quantum 3WI has been considered in three forms.
In one, all particles are bosons, in the ``Fermion I'' model 
two fermions interact to form a boson, and in the 
``Fermion II'' model a boson and a fermion interact to form a fermion.
In this paper we consider the Fermion I model.
We find the associated first quantization problem,
obtain the eigenfunctions of the Hamiltonian in closed form,
and derive the Bethe equations rigorously from the requirement
that the wave functions be periodic.  

Because the model is first
order in the derivatives, the energy is not bounded below.
Nevertheless, as in Dirac's model of the electron, a stable ground
state may be constructed by filling all negative energy states.
One can then consider all states with reference to the filled
Dirac sea of negative energy states.  
This approach is used in Korepin$^3$ and in Bergknoff and 
Thacker$^4$ to calculate the excitation
spectrum of the massive Thirring model from the Bethe ansatz states.

We prove
the existence of a thermodynamic limit as the length of
the interval tends to infinity with constant density.  The Dirac sea is
obtained in the infinite density limit.  Finite particle/hole
excitations from the Dirac sea correspond to the solutions of
certain integral equations.  Such excitations have positive relative energy,
and we determine the asymptotic relation between energy and momentum.
Following the treatment of a one-dimensional boson gas with
delta-function interactions by Yang and Yang$^5$, 
we derive a functional for the 
relative free energy of finite (local) density perturbations from
the Dirac sea.  We prove that the functional is convex and bounded below
and obtain the equation for thermal equilibrium.

Kulish and Reshetikhin$^2$ used the quantum inverse scattering method 
(also called the algebraic Bethe Ansatz) to investigate $GL(N)$ invariant 
transfer matrices on a one dimensional lattice.  They obtained quantized 
$N$ wave models as a formal continuum limit and found explicitly the Bethe
equations for the bosonic 3WI.  Wadati and 
Ohkuma$^1$ also considered the bosonic case.  Ohkuma and Wadati$^6$
treated the Fermion I and Fermion II cases and gave an
algorithm for constructing the Bethe wave functions.

Ohkuma$^7$ investigated the Fermion I quantization of the three wave
interaction, wrote down the Bethe equations, and calculated
the free energy and excitations from a ground state.
However a momentum cut-off is introduced in Ohkuka$^7$.  The results depend 
on the cut-off and do not have a finite limit when the cut-off is
taken to infinity.  
By constructing all physical states with reference to a Dirac sea we
obtain that are finite and are independent of any cut-off.  
The methods provide a basis for
calculating the free energy of other integrable models whose 
spectrum is unbounded below.  

The quantized field of the 3WI may be rewritten as a 
first quantization for a many body problem. Such a reduction 
was carried out by Lieb and Liniger$^8$ for the 
quantization of a Bose gas -- essentially the
quantization of the nonlinear Schr\"odinger equation.
They obtained a Helmholtz equation in the interior of 
the fundamental domain, together with boundary conditions
at the interfaces associated with the interactions; see also
Korepin, Bogoliubov, and Izergin$^9$.
A novel feature of the 3WI is that
the system of equations links functions with different
numbers of variables, corresponding to the fact that individual 
particle numbers are not conserved. 

As in Lieb and Liniger$^8$ we solve the differential equations in 
certain fundamental cones
$\Cal C^{m,n,p}$ and then extend the densities to all space, first by
anti-symmetrizing across the boundaries of $\Cal C^{m,n,p}$ and then by
periodicity.   
The differential equations for the densities force 
discontinuities where fermions of different types interact.
However like fermions do not interact, so to avoid 
discontinuities on the boundaries of the fundamental cones
we impose a regularity requirement, (2.5).  This requirement
forces one to anti-symmetrize the basic 
solutions of the differential equations over momenta.
This in turn is fundamental to obtaining the Bethe equations, 
as we show by example in \S5.

In \S\S3,4 we sketch, in greater detail than in Wadati
and Ohkuma$^1$, Wadati and Ohkuma$^6$, a
derivation of an algorithm for the eigenfunctions of the 3WI,
and go on to obtain the explicit form of the eigenfunctions 
(cf. Theorem 5.1).  This form 
generalizes the Bethe Ansatz (Bethe$^{10}$, Lieb and Liniger$^8$, Lieb$^{11}$, 
Yang and Yang$^{12}$) to
systems in which the eigenvectors are superpositions of states having
different numbers of particles.  In \S5 we derive the Bethe equations 
from the requirement of periodicity.

In \S6 we reformulate the finite Bethe equations as a system of
integral equations and prove existence of a certain
thermodynamic limit at finite density. The Dirac sea is then obtained
as an infinite density limit. In \S7 integral equations for the
particle/hole excited states are obtained, and energy/momentum
asymptotics are found.
Finite density perturbations of the Dirac sea are constructed in
\S8.  In \S9 we investigate the (relative) free energy functional.
The formal Euler-Lagrange equations for its minimum are the same as
those proposed by Ohkuma$^7$ if one imposes a cut-off.
Without a cut-off they have no strict
solutions, but there is a natural interpretation
which does yield meaningful equations for the thermal equilibrium
state.

\ms\ni
{\bf 2. THE THREE WAVE INTERACTION}
\ss

The Hamiltonian of the classical three-wave interaction is 
$$
\align
H =& \int  
\sum_{j=0}^2 i c_j \frac{\partial \Psid_j}{\partial x}\Psi_j 
 + \g( \Psid_0 \Psi_1 \Psi_2 + \Psid_2 \Psid_1
	\Psi_0 )\, dx  \tag2.1\\
	 =& T + \g (V^{\dagger} + V),
\endalign
$$ 
where $\Psi_j$ are complex valued functions and $\dagger$ denotes complex 
conjugation. The constants $c_j$ are wave speeds, and $\g$ is a coupling 
constant. 
Zakharov and Manakov$^{13}$ found a Lax pair for the classical system
and Beals and Sattinger$^{14}$ gave a set of action-angle variables.
However, integrability of the classical system will 
play no role in our analysis of the quantized model. 

The system is formally quantized by taking the $\Psid_j$, $\Psi_j$ to be 
creation and annihilation operators on a Fock space, $\dagger$ denoting 
Hermitian adjoints.  We take 
$\Psi_0(x,t)$ to be a boson field operator, and $\Psi_1,\ \Psi_2$ to be  
fermion fields. Thus the operators are required to satisfy the equal time 
commutation ($[\ ,\ ]_-$) or anti-commutation  ($[\ ,\ ]_+$) relations
$$
\align
[ \Psi_j(x,t),\Psi_{j'}(x',t)]_+=&0,\quad 
[ \Psi_j(x,t),\Psid_{j'}(x',t)]_+=\delta_{jj'}\delta(x-x'), 
\\
[ \Psi_0(x,t),\Psi_0(x',t)]_-=&0,\quad
[ \Psi_0(x,t),\Psid_0(x',t)]_-=\delta(x-x'), \tag2.2 \\
[\Psi_0(x,t),\Psi_j(x',t)]_-=&0,\quad
[\Psid_0(x,t),\Psi_j(x',t)]_-=0,\quad j,j'=1,2.
\endalign
$$

The Hermitian operator $T$
represents the kinetic energy, with distinct wavespeeds
$c_j$ for each field.  The interaction
term $V^{\dagger}$ formally anihilates the two particles with wavespeeds 
$c_1$ and
$c_2$ (particles of type 1 and 2) and creates one with wavespeed
$c_0$, while $V$ destroys a type 0 particle and creates types
1 and 2.

We assume throughout that the speeds of the fermions bracket the boson 
speed: $(c_1-c_0)(c_2-c_0)<0$. In addition, we make the simplifying 
assumption that
$$
c_1-c_0=\alpha=c_0-c_2 \tag2.3
$$
where $\alpha$ is an arbitrary constant. Without loss of generality we
assume throughout this paper that $\alpha>0$ 

The Hamiltonian is defined on a Hilbert space $\Cal H$ consisting of the tensor
product of a boson and two fermion Fock spaces.
There exists a non-zero vector $|0\rangle$ in the Fock space, called the 
pseudovacuum, such that
$$
\Psi_j(x,t)|0\rangle = 0, \qquad j=0,1,2.
$$
Define particle number operators
$$
N_j = \int dx\, \Psid_j \Psi_j, \qquad j=0,1,2.   
$$

The coordinates of particles of types 0,1, and 2 are denoted by 
$x_j,\, y_j,\, z_j$ respectively. Let 
$$
\Psid_0(x)=\Psid_0(x_1,\dots, x_m)=\Psid_0(x_1)\cdots\Psid_0(x_m)
$$
and define $\Psid_1(y)$ and $\Psid_2(z)$ similarly. Then set
$$
\Psid(x,y,z)=\Psid_0(x)\Psid_1(y)\Psid_2(z),
\qquad
x\in  R^m,
y\in
 R^n,\
z\in R^p.
$$
Fundamental states in $\Cal H$ are given formally by
$$
\Psid(x,y,z) |0\rangle,
$$
which are eigenstates of $N_0$,$N_1$, and $N_2$ with 
eigenvalues $m$,$n$, and $p$ respectively.  
A general state vector in this subspace may be written in the form
$$
|\psi_{m,n,p}\rangle=\frac{1}{m!n!p!}\int_{R^{m,n,p}}
f_{mnp}(x,y,z)\Psid(x,y,z) |0\rangle dx\,dy\,dz, \tag2.4
$$
where $f_{mnp}$ belongs to $\Cal L_2(R^{m,n,p})$
 and $ dx = dx_1 \,dx_2\, \dots \,dx_m $,
{\it etc.}
Since $\Psid(x,y,z)$ is symmetric in the $x_j$ and anti-symmetric in the 
$y_k$ and $z_l$, the density function $f_{mnp}$ may be assumed to possess 
these same symmetries.

If a density $f_{mnp}$ is associated with an eigenstate, then, as
shown in \S4, it must be discontinuous across the hyperplanes
$y_k=z_l$, which we call the {\it interaction manifolds\/}.
It is natural to assume that these are the only discontinuities.
We therefore assume continuity across the hyperplanes
$y_k=y_{k+1}$, provided $y_k\ne z_l$ for all $z_l$, and similarly with
respect to the $z$ coordinates.  Continuity plus anti-symmetry 
across these hyperplanes implies:
$$
\align &Regularity\ condition:\ f_{mnp} \text{ vanishes as}\  
y_k\tt y_{k\pm 1}\ \text{ if} \\ 
&\qquad y_{k\pm 1}\ne z_l\ \text{for all}\ l
\text{ and similarly as} \ z_l\tt z_{l\pm 1}\ne y_k,\ \text{ all}\ 
k. \tag2.5
\endalign
$$
By symmetry, we  may replace the integration in (2.4) by integration over 
the cone $\Cal C^{m,n,p}$ given by
$$
x_1<x_2<\dots < x_m;\ \ y_1<y_2<\dots < y_n;\ \ z_1<z_2<\dots < z_p,
$$
and write 
$$
|\psi_{m,n,p}\rangle=\int_{\Cal C^{m,n,p}}
f_{mnp}(x,y,z)\Psid(x,y,z) |0\rangle dx\,dy\,dz. \tag2.6
$$

{From} now on, we write the state vectors in the form (2.6) and assume that
the densities $f_{mnp}$ are supported on the cone $\Cal C^{m,n,p}$
and satisfy the regularity condition (2.5), which implies that 
the antisymmetric extensions of the $f_{mnp}$ are
continuous across the collision manifolds $y_j=y_k$ and $z_j=z_k$ 
of like fermions.

The classical version of the Hamiltonian (2.1) is integrable and possesses 
an infinite number of conservation laws: Zakharov and Manakov$^{13}$. 
Three important conserved quantities for each version are
$$
\align
N  =& \int 2\Psid_0 \Psi_0 + \Psid_1 \Psi_1 + \Psid_2
\Psi_2\, dx 
\\ 
D  =& \int  \Psid_1 \Psi_1 - \Psid_2 \Psi_2 \, dx
\\ 
P  =& \int  
\sum_{j=0}^2 i \Psid_j \frac{\partial \Psi_j}{\partial x} 
\,dx
\endalign
$$
In the quantized version operators $N$ and $D$ give 
conserved particle
numbers, and $P$ is the momentum operator. 
Indeed it is not difficult to show from (2.2)
that $N,D,P$ commute with $H$.

\ms\ni
{\bf 3. REDUCTION TO FIRST QUANTIZATION}
\ss

The quantum field problem can be rewritten as a first quantization problem 
for a many body problem.

Denote the subspace of states $|\psi_{m,n,p}\rangle$ by ${\Cal F}_{mnp}$. 
These are eigenstates of $N$ and $D$ with eigenvalues
$N=2m+n+p$ and $D=n-p$ respectively, but are not eigenstates of the 
Hamiltonian $H$, since $V^{\dagger}$ and $V$ create and destroy particles. 
However, since $N$ and $D$ commute with $H$, the quantum numbers $N$ and 
$D$ label a subspace that is invariant under $H$, $N$, and $D$. 
We denote this space by $\Cal F_{N,D}$. 

We now fix $N$ and $D$.  Within  $\Cal F_{N,D}$,
densities can be labeled by $m$ alone and we denote the 
corresponding subspace by  $\Cal F_m$. 
Note that both of $N\pm D$ must be non-negative even integers. 
When there are no bosons 
present, ($m=0$), there are $(N\pm D)/2$ fermions of types 1 and 2 
respectively. The maximum number of bosons occurs when the maximal
number of fermions have combined, yielding
$M=\min\, (N\pm D)/2=\tfrac12(N-|D|)$. 

The operators $T,V^{\dagger},$ and $V$ act on the subspaces $\Cal F_m$ 
as follows:
$$
T :\Cal F_m \rightarrow \Cal F_m, 
\qquad 
V^{\dagger}: \Cal F_m \rightarrow \Cal F_{m+1}, 
\qquad 
V : \Cal F_m\rightarrow \Cal F_{m-1}.
$$
An eigenstate of the Hamiltonian in $\Cal F_{N,D}$ 
is a sum of vectors $|\psi_m\rangle\in\Cal F_m$,
and the eigenvalue equation becomes the coupled system 
$$
\align
(E-T)|\psi_0\rangle =&\g V |\psi_1\rangle\\
\vdots \\
(E-T) |\psi_m\rangle = &\g V^{\dagger}|\psi_{m-1}\rangle+\g V |\psi_{m+1}\rangle \tag3.1\\
\vdots  \\
(E-T)|\psi_M\rangle=&\g V^{\dagger}|\psi_{M-1}\rangle
\endalign          
$$
since $V|\psi_0\rangle=V^{\dagger}|\psi_M\rangle=0$.

Let $f_m$ be the density corresponding to the vector $|\psi_m\rangle$.
Equations (3.1) may be converted to a system of partial differential 
equations by calculating the action of $T$, $V^{\dagger}$ and $V$ on the 
densities $f_m$. The calculation of the action of $T$ is straightforward, 
while the calculation of $V$ and $V^\dagger$ is somewhat more involved; 
details are given in Beals, Sattinger, and Williams$^{15}$.  
The action of $T$ is:
$$
\align
T|\psi_m\rangle=&T\int_{ \Cal C^{m,n,p}}
f_m(x,y,z)\Psid(x,y,z)|0\rangle\,dx\,dy\,dz \\
=&\int_{ \Cal C^{m,n,p}}
\left(X_mf_m(x,y,z)\right)\Psid(x,y,z)|0\rangle\,dx\,dy\,dz 
\endalign
$$
where
$$
X_mf=ic_0\sum_{j=0}^m \frac{\partial f}{\partial x_j}
+ic_1\sum_{k=0}^n \frac{\partial f}{\partial y_k}
+ic_2\sum_{l=0}^p \frac{\partial f}{\partial z_l}. \tag3.4
$$
We use $V^{\dagger}$ to denote the action on densities as well as the
action on vectors.  Then
$$
V^{\dagger}f_m= \sum_{j,k,l=1}^{m+1,n,p}
(-1)^{n(m)+k+l}f_m(\hat x_j; \langle y,x_j,k\rangle;
\langle z, x_j,l \rangle ), \tag3.5
$$
where  $(x,y,z)\in \Cal C^{m+1,n-1,p-1}$ and
$$\align
\hat x_j&=(x_1, \dots,x_{j-1},x_{j+1}, \dots x_m) \in  \Cal C^m,\\
\langle y,x_j,k\rangle&=(y_1,\dots,y_{k-1},x_j,y_k,\dots,y_{n-1})\in  
\Cal C^n,
\ etc.\endalign
$$

\demo{Remarks} 1. Since $f_m$ is supported on $\Cal C^{m,n,p}$,
$f_m(\hat x_j; \langle y,x_j,k\rangle ; \langle z,x_j,l\rangle)=0$
unless $x_j\in(y_{k-1},y_k)\cap (z_{l-1},z_l)$.  Hence
$$
(V^{\dagger}f_m)(x;y;z)= 
\sum_{j=1}^{m+1}(-1)^{n(m)+k_j+l_j}f_m (\hat x_j; \langle y ,x_j,k_j\rangle;
\langle z, x_j,l_j \rangle ), \tag3.6
$$
where $k_j$ and $l_j$ are uniquely determined by the constraint
$$
x_j\in (y_{k_j-1},y_{k_j})\cap (z_{l_j-1},z_{l_j}). \tag3.7
$$

2. In (3.6) we interpret the value of $f_m$ on the interaction manifold
$x_j=y_{k_j}=z_{l_j}$ as the average of its values as $y_{k_j}\tt z_{l_j} \pm$.

3.  The action of $V^{\dagger}$ can be thought of as
gluing a type 1 fermion at $y_k$ and a type 2 fermion at $z_l$
to form a boson at $x_j$.  The coordinates are relabeled to take into
account the shift in numbers of variables, summed over all possible pairs. 
Similarly, the action of $V$ can be thought of as the
decomposition of a boson into a pair of fermions, summed over the bosons.
In fact the operation of $V$ on densities is given by
$$
(Vf_m)(x,y,z)=\sum_{k,l=1}^{n+1,p+1} (-1)^{n(m)+k+l}
f_m (\langle x,y_k,j_k \rangle;\hat y_k; \hat z_l) \delta(y_k-z_l). \tag3.9
$$
where $j_k$ is uniquely determined by the constraint $y_k\in
(x_{j_k-1},x_{j_k})$.
\enddemo

We illustrate in some low order cases.  For $2m+n+p=N=2, \ 
f_0=f_0(y,z),\ f_1=f_1(x)$:
$$
\align
(V^{\dagger}f_0)(x)=&-f_0(x,x),\quad  Vf_0=0\\
V^{\dagger}f_1=&0,\qquad\quad (Vf_1)(y,z)=f_1(y)\delta (y-z)
\endalign
$$
\ss
For $ 2m+n+p=N=4,\ f_0=f_0(y_1,y_2,z_1,z_2), \ f_1=f_1(x,y,z),\ f_2=
f_2(x_1,x_2)$:
$$
\align
(V^{\dagger}f_0)(x,y,z)=&f_0(x,y,x,z)-f_0(x,y,z,x)-f_0(y,x,x,z)+f_0(y,x,z,x),\\
(V^{\dagger}f_1)(x_1,x_2)=&-f_1(x_1,x_2,x_2)-f_1(x_2,x_1,x_1),\\
\qquad V^{\dagger}f_2=&0;\\
Vf_0=&0 \\
(Vf_1)(y_1,y_2,z_1,z_2)=&f_1(y_1,y_2,z_1)\delta(y_1-z_2)
-f_1(y_1,y_2,z_2)\delta(y_1-z_1),   \\
&-f_1(y_2,y_1,z_1)\delta(y_2-z_2)+f_1(y_2,y_1,z_2)
\delta(y_2-z_1),\\
(Vf_2)(x,y,z)=&[f_2(y,x)+f_2(x,y)]\delta(y-z).
\endalign
$$

\ss\ni
{\bf 4. THE BETHE ANSATZ}
\ss

The differential equations for the densities, supplemented with the 
regularity condition (2.5) at the boundaries of the cones $\Cal C^{m,n,p}$, 
describe the spectral problem in first quantization.

{From} the results of \S3 the differential equations for the densities are 
$$
\gather
X_0f_0+\g Vf_1=Ef_0,\\
\vdots  \\
X_mf_m+\g Vf_{m+1}+\g V^{\dagger}f_{m-1}=Ef_m \tag4.1 \\
\vdots \\
X_Mf_M+\g V^{\dagger}f_{M-1}=Ef_M
\endgather
$$

Wadati and Ohkuma$^1$ gave an algorithm for constructing the general 
eigenstates.  In the next two sections 
we find these eigenstates in closed form, in two steps. 
In this section we 
construct ``local solutions'' of equations (4.1);
on each of the cones $\Cal C^{mnp}$ is a multiple of an
exponential solution of the associated differential equation.

Because the operator
$V$ introduces delta function singularities on the interaction manifolds, 
solutions must have discontinuities
across these manifolds.  
However the local solutions violate (2.5) and thus they 
have additional discontuities if
extended by anti-symmetry.  We denote the densities
of the local solutions by $h_m$ to distinguish them from the regular 
densities, which will satisfy (2.5).  We obtain these regular densities
in \S5 by taking
linear combinations of the $h_m$: anti-symmetrizing over the momenta.  

\ss\ni
{\bf Low Order Cases} 
\ss
Two low order cases illustrate the main features of
these equations.  For $N=2,\ D=0$, equations (4.1) are
$$
\align
ic_1\frac{\partial h_0}{\partial y}+ic_2\frac{\partial h_0}{\partial z}+
 \g h_1(y)\delta(y-z)=&Eh_0(y,z) \tag4.2a \\
ic_0\frac{\partial h_1 }{\partial x}- \g h_0(x,x)=&Eh_1(x) \tag4.2b
\endalign
$$
where $h_0=h_0(y,z),\ h_1=h_1(x)$.  In the first equation the interaction 
term vanishes off the interaction manifold $y=z$ so we take the 
solutions to be multiples of an exponential in each of the open sectors
$y<z$ and $y>z$ of $\Cal C^{0,1,1}$:
$$
h_0(y,z)=\cases \phi_1^0\,\exp{i(\eta y-\zeta z)}, & y<z, \\
 \phi_2^0\,\exp{i(\eta y-\zeta z)}, & y>z,
\endcases
$$
where $E=-(c_1\eta-c_2 \zeta)$, and $\phi_1^0$ and $\phi_2^0$ are constants 
to be determined.

In (4.2b) we take $h_0(x,x)$ on the interaction manifold to be the average 
of the values of $h_0(y,z)$ {from} the left and right.
Hence the second equation is
$$
\left(ic_0\frac{\partial }{\partial x}-E\right)h_1(x)=
\g\, \frac{\phi_1^0+\phi_2^0}{2}\,e^{i(\eta-\zeta)x}.
$$

We seek a solution of the form $h_1(x)=
Ae^{i(\eta-\zeta)x}$, so that
$$
h_1(x)=g(\eta,\zeta)\,\frac{\phi_1^0+\phi_2^0}{2}\,\exp{i(\eta-\zeta)x},
\qquad
g(\eta,\zeta)=\frac{\g}{(c_1-c_0)\eta-(c_2-c_0)\zeta}.
$$
By (2.3),
$$
g(\eta ,\zeta)=\frac{\g}{\alpha(\eta+\zeta)}:=g(\eta+\zeta).
$$
We use this form of $g$ {from} now on.

Finally, we determine the ratio $\phi_1^0/\phi_2^0$ by integrating (4.2a) 
across the interaction manifold $y=z$.  We change variables, letting 
$d=y-z,\ s=y+z$, and obtain
$$
\align
i(c_1-c_2)&\frac{\partial h_0}{\partial d}+i(c_1+c_2)
\frac{\partial h_0}{\partial s}+\g h_1\delta (d)
=Eh_0,\\
h_0=&h_0\left(\frac{s+d}{2},\frac{s-d}{2}\right),
\qquad
h_1=h_1\left(\frac{s+d}{2}\right)
\endalign
$$
Integrating across the jump at $d=0$ ($y=z=x$) we get
$$
i(c_1-c_2)(\phi_2^0-\phi_1^0)\exp ix(\eta-\zeta)+\g g(\eta+\zeta)
\frac{\phi_1^0+\phi_2^0}{2}\exp ix(\eta-\zeta)=0;
$$
hence
$$
\frac{\phi_2^0}{\phi_1^0}=\frac{2i(c_1-c_2)-\g g}{2i(c_1-c_2)+\g g}.
$$
Using our assumption (2.3) we may write this in the simplified form
$$
\phi_2^0=\theta(\eta+\zeta)\phi_1^0,
\qquad
\theta (\l)=\frac{\l+i\omega}{\l-i\omega}, 
$$
where
$$
\omega=\frac{\g^2}{4 \alpha^2}.
$$
Setting $\phi^0_1=1$, we have
$$
h_0(y,z)=\cases \exp i(\eta y-\zeta z), & y<z;\\
\theta(\eta-\zeta)\,\exp i(\eta y-\zeta z), & y>z;
\endcases
$$
and
$$
h_1(x)=\tfrac12 g(\eta+\zeta)\,(1+\theta(\eta+\zeta))\,\exp i x(\eta-\zeta).
$$
\ss

For $N=3,\ D=1$ equations (4.1) are
$$
\align
ic_1\frac{\partial h_0}{\partial y_1}&+ic_1\frac{\partial h_0}{\partial y_2}
+ic_2\frac{\partial h_0}{\partial z}-Eh_0(y_1,y_2;z)\\
=& \g (h_1(y_1;y_2)\delta(y_1-z)-h_1(y_2;y_1)\delta(y_2-z)) \tag4.3a\\
ic_0\frac{\partial h_1}{\partial x}&+ic_1\frac{\partial h_1}{\partial y}
-Eh_1(x;y)=\g (h_0(y,x;x)-h_0(x,y;x)), \tag4.3b
\endalign
$$
with $h_0=h_0(y_1,y_2;z)$ and $h_1=h_1(x;y)$.
Arguing as above, we obtain 
$$
h_0(y;z)=\cases \exp i(\eta_1y_1+\eta_2y_2-\zeta z), & y_1<y_2<z; \\
\theta (\eta_2+\zeta)\,\exp i(\eta_1y_1+\eta_2y_2-\zeta z), & y_1<z<y_2;\\ 
\theta(\eta_1+\zeta)\theta (\eta_2+\zeta)\,\exp i(\eta_1y_1+\eta_2y_2-\zeta z), 
& z<y_1<y_2;
\endcases
$$
and
$$
h_1(x;y)=\frac{1}{2}\cases \left(1+\theta(\eta_2+\zeta)\right)
g(\eta_2+\zeta)\,\exp\, i((\eta_2-\zeta)x+\eta_1y) & y<x;\\
-\theta(\eta_2+\zeta)(1+\theta(\eta_1+\zeta))
g(\eta_1+\zeta)\,\exp\, i((\eta_1-\zeta)x+\eta_2 y) & x<y. 
\endcases
$$

\ni
{\bf The General Case}
\ss

The two preceding cases show that when $y_k$ and $z_l$ are transposed, 
(with $y_k$ moving to the right of $z_l$), $h_0$ is multiplied by 
$\theta(\eta_k+\zeta_l)$, where $\eta_k, -\zeta_l$ are the momenta of
the corresponding fermions. Since the differential equations are local 
and involve only pairwise interactions, this situation extends to the general
case.

We begin by introducing some notation and conventions.
For fixed $N$ and $D$, $\Cal C^{m,n,p}$ is uniquely determined by $m$, so 
{from} now on we denote it by $\Cal C^m$. The domain $\Cal C^m$ is a union of 
$$
\frac {(m+n+p)!}{m!n!p!}
$$
open sectors determined by the interspacing of the particles of type 0,1,2. 
We label them by $m+n+p$-tuples $\s=(\e_1,\e_2,\dots, \e_{m+n+p})$
where $\e_j=0,-1$, or $+1$ if the particle in the $j^{th}$ position 
along the real line is of type 0,1, or 2 respectively.  For example
$$
\s=(1,0,-1,1) \leftrightarrow \ \{z_1<x_1<y_1<z_2\}.
$$
The boundary of $\Cal C^m$ consists  of the interaction manifolds $y_j=z_k$
together with the sets $y_j=y_{j+1}$ and $z_k=z_{k+1}$ at which like fermions
coalesce.

 A {\it transposition} consists of interchanging a -1 with a +1 to its 
immediate right, {\it e.g.} $(1,0,-1,1) \mapsto (1,0,1,-1)$. Each 
transposition corresponds to interchanging an adjacent pair of coordinates 
$y_k,\,z_l$; thus to every interaction manifold there is a corresponding 
transposition. The sectors of $\Cal C^m$ are partially ordered by the number 
of transpositions; thus, for example $(1,0,-1,1)<(1,0,1,-1)$. By definition, 
no transpositions occur between -1's and 1's separated by one or more zeroes.

Since $Vh_1$ is supported on the interaction manifold, $h_0=h_0(y;z)$ 
satisfies 
$$
X_0h_0=ic_1\sum_{j=1}^n\frac{\partial h_0}{\partial y_j}
+ic_2\sum_{k=1}^p\frac{\partial h_0}{\partial z_k}=Eh_0 \tag4.4
$$
in each open sector $\s \subset \Cal C^0$.

Let
$$
\s_0=\{y_1<\dots<y_n<z_1<\dots<z_p\} \sim (-1,\dots,-1,1,\dots,1).
$$
We take
$$
h_0=\exp i\left(\sum_{j=1}^n \eta_j y_j-\sum_{k=1}^p\zeta_k z_k\right),
\qquad
(y;z)\in \s_0
$$
with 
$$
E=-c_1\sum_{j=1}^n\eta_j+c_2\sum_{k=1}^p\zeta_k. \tag4.5
$$
The solution in a general sector $\s\in \Cal C^0$ is
$$
h_0=\phi^0_{\s}\exp i\left(\sum_{j=1}^n \eta_j y_j-\sum_{k=1}^p\zeta_k 
z_k\right),\qquad y;z\in\s, \tag4.6
$$
where the constant $\phi^0_{\s}$ is uniquely determined by $\s$, and is 
given in (4.8) below.

\proclaim{Lemma 4.1}  Let $\s<\s'$  differ by a single transposition of 
fermions of types 1 and 2 at $y_j,\,z_k$ with momenta $\eta_j$ and 
$-\zeta_k$ respectively.  Then
$$
\phi^0_{\s'}=\theta (\eta_j+\zeta_k)\phi^0_{\s},  \tag4.7
$$
\endproclaim

\demo{Remark} For $\s'<\s$ the phase factor is 
$$
[\theta(\eta_j+\zeta_k)]^{-1}=\theta(-\eta_j-\zeta_k).
$$
Thus within a {\it fixed\/} particle number sector these conditions
on the wave function are of the usual Bethe Ansatz type; the ratio of
phases that differ by a permutation of neighboring particles is a
function of the momenta.\enddemo

\demo{Proof} All the interactions are pairwise and local, so the same 
computations for the case $N=2,\ D=0$ extend to the general case.
The ratio of the constants $\phi^0_{\s},\ \phi^0_{\s'}$ is determined 
entirely {from} the first and second equations in the hierarchy (4.1).\qed
\enddemo

Each $\s\in \Cal C^0$ is obtained {from} $\s_0=(-1,\dots,-1,1,\dots,1)$
by a unique set of transpositions $(y_j,z_k)$, which we denote by 
$T_{\s}$.  Thus
$$
\phi^0_{\s}=\prod_{(j,k)\in T_{\s}}\theta(\eta_j+\zeta_k). \tag4.8
$$

Now consider the higher order components of the density: $h_m$, $m\ge 1$.
The sectors of $\Cal C^m$ are characterized by $(m+n+p)$-tuples $\s$ with 
$m$ zeroes.  We assign momenta to $\s$  as follows. 
Regard each 0 in $\s$ as a pair $(-1,1)$.  Starting {from} the left, assign 
the  -1's momenta $\eta_1, \eta_2,\ \dots$; the +1's momenta 
$-\z_1,-\z_2,\dots$; and assign the 0's the sums of momenta $\eta_j-\zeta_k$, 
with the $\eta$'s and $\zeta$'s taken in their normal order. For example,
$$
\gather
(-1,0,1,0) \leftrightarrow (\eta_1, \eta_2-\z_1,-\z_2,\eta_3-\z_3);\\
(-1,1,1,0,1,0,-1) \leftrightarrow (\eta_1,-\zeta_1,-\zeta_2,\eta_2-\zeta_3,
-\zeta_4,\eta_3-\zeta_5,\eta_4).
\endgather
$$

For fixed locations of the zeroes in $\s$ the basic exponential solution is 
defined by
$$
e^m_
{\s}=\exp i \left( \sum_{r=1}^m x_r(\eta_{i_r}-\z_{j_r} )+
\sum_{s=1}^n y_s\eta_{a_s}-\sum_{t=1}^p z_t \z_{b_t} \right) \tag4.9
$$
where                          
$$
\align
i_r\ (j_r)=&\ \text{\# $x$'s and $y$'s ($z$'s) to the left of $x_r$, 
inclusive of $x_r$};\tag4.10\\
a_s\ (b_t)=&\ \text{\# $x$'s and $y$'s ($z$'s) to the left of $y_s$, ($z_t)$, 
inclusive of $y_s$, ($z_t$)}.
\endalign
$$
\demo{Remark} {From} (3.7), $i_r=r+k_r-1,\ j_r=r+l_r-1.$\enddemo

For example    
$$
\align
e^m_{\s}(x;y;z)=&\exp ix_1
\left( x_1(\eta_2-\z_1)+x_2(\eta_3-\z_3)+y_1\eta_1-z_1\z_2\right),\\
&\qquad \s=(-1,0,1,0);\\
e^m_{\s}(x;y;z)=&\exp i
\left(x_1(\eta_2-\z_3)+x_2(\eta_3-\z_5)+y_1\eta_1 +y_2\eta_4-z_1\z_1
-z_2\z_2-z_3\z_4\right)\\
&\s=(-1,1,1,0,1,0,-1)
\endalign
$$

We now describe the general case. For any $\s\in \Cal C^m$ define $g_{\s}$ by
$$
g_{\s}=\prod_{r=1}^m g_{i_r j_r},
\qquad
g_{i_r j_r}=g(\eta_{i_r}+\zeta_{j_r}),  \tag4.12
$$
where $i_r,\ j_r$ are given in (4.10).
For example, for $(-1,0,1,0)$, $i_1=2,\ j_1=1,\ i_2=3,\ j_2=2$, and
$$
g_{\s}=g_{21}g_{32}.
$$

The components of $\Cal C^m$ are determined by the location of the zeroes and
the numbers and orders of $+1$'s and $-1$'s between successive zeroes. The 
factor $g_{\s}$ is constant on each connected component of $\Cal C^m$.
For $\s\in \Cal C^m$, we define
$$
\phi^m_{\s}=\phi^0_{\Sigma}\prod_{r=1}^m (-1)^{n(m)+k_r+l_r}
\frac{1+\theta(\eta_{i_r}+\zeta_{j_r})}{2}, \tag4.13
$$
where $\Sigma\in \Cal C^0$ is obtained {from} $\s$ by replacing each 
$0$ in $\s$ by
$(-1,1)$, and $\phi^0_{\Sigma}$ is given by (4.8).

\proclaim{Theorem 4.2}The basic solutions of equations (4.1) are given by
$$
h_m(x,y,z)=\psi_{\s}^m e^m_{\s}(x,y,z), \tag4.14
$$
where 
$$
\psi_{\s}^m =\phi_{\s}^m g_{\s}
=\phi^0_{\Sigma}\left(\frac{\g}{\alpha}\right)^m\prod_{r=1}^m 
\frac{(-1)^{n(m)+k_r+l_r}}{\eta_{i_r}+\zeta_{j_r}-i\omega}
$$
and the $\phi_{\s}^m$ are given by (4.8), (4.13); the $g_{\s}$  by (4.12);
and the $e^m_{\s}$ by (4.9).
\endproclaim
The proof is given in Beals, Sattinger, and Williams$^{15}$.  
In addition to Lemma 4.1, it
relies on the identity
$$
(-1)^{n(m)+k_r+l_r}\,\frac{\phi^m_{\sigma_r}+\phi^m_{\sigma'_r}}{2}
=\phi^{m+1}_\sigma, \qquad 1\le r\le m+1
$$
where $\sigma_r$ and $\sigma'_r$ are obtained from $\sigma$
by replacing the $r$-th zero by $(-1,1)$ and by $(1,-1)$
respectively.  This condition insures compatibility between the
phases in {\it different\/} particle number sectors; it has
no analogue in the usual Bethe Ansatz.

\ms\ni
{\bf 5. THE BETHE EQUATIONS}
\ss
In \S 4 we constructed basic solutions $h_m$ of equations (4.1).  
As noted in \S2 we want to
impose the regularity condition (2.5) that eigendensities
have no unnecessary discontinuities, i.e.  
that their anti-symmetric extensions be continuous across the collision 
manifold of like fermions. To obtain this property, 
we antisymmetrize the basic solutions over the momenta. 

Permuting the momenta of the basic solution of (4.4), 
(4.5) gives another solution. The general solution of (4.4, 4.5) is 
a linear combination
$$
e_0(J,K)=\exp i\left(\sum_{j=1}^n y_j\eta_{J_j}
-\sum_{k=1}^p z_k\zeta_{K_k}\right),  
$$
where $J,\,K$ are permutations of $1, \dots,n$ and 
$1,\dots,p$ respectively. In particular,
$$
f_0(y;z)=\sum_{J,K}\text{sgn}(J)\,\text{sgn}(K)\,e_0(J,K), 
$$
satisfies $X_0f_0=Ef_0$ in 
$$
\s_0=\{y_1<\dots<y_n<z_1<\dots<z_p\},
$$
and satisfies the regularity condition (2.5).

The density in the other sectors is given by
$$
f_0(y;z)=\sum_{J,K}\text{sgn}(J)\,\text{sgn}(K)\, 
\phi^0_{\s}(J,K)\,e_0(J,K), \tag5.1
$$
where $(y;z)\in \s,$ and
$$
\phi^0_{\s}(J,K)=\prod_{(j,k)\in T_{\s}}\theta(\eta_{J_j}+\zeta_{K_k}).
$$

The corresponding densities for higher $m$ are consequently
$$
\align
f_m(x;y;z)=&\sum_{J,K}\text{sgn}(J)\,\text{sgn}(K)\, 
\phi^m_{\s}(J,K)\,g_{\s}(J,K)\,e^m_{\s}(J,K),
 \tag5.2\\
&(x;y;z)\in \s,
\endalign
$$
where $e^m_{\s}(J,K)$, $g_{\s}(J,K)$ and $\phi^m_{\s}(J,K)$ are obtained 
{from} (4.9), (4.12), and (4.13) by making the substitutions 
$j,\ k \mapsto J_j,\ K_k.$
For example, to obtain densities which vanish when $y_1=y_2\ne z$ in the
case $N=3,\ D=1$, we antisymmetrize the base function over the momenta: setting
$$
e_0(j,k)=\exp i(\eta_{j}y_1+ \eta_k y_2-\zeta z),
$$
we take
$$
f_0(y_1,y_2,z)= 
\cases e_0(1,2)-e_0(2,1), & y_1<y_2<z ;\\
	\theta(\eta_2+\zeta)\,e_0(1,2)-\theta(\eta_1+\zeta)\,
e_0(2,1), & y_1<z<y_2; \\
\theta(\eta_1+\zeta)\theta(\eta_2+\zeta)\,[e_0(1,2)-e_0(2,1)], & z<y_1<y_2.
\endcases
$$
Now $f_0(y_1,y_2,z)$ vanishes when $y_1=y_2 \ne z$; but we can not assert 
anything for the case $y_1<z<y_2$, since $y_1$ and $y_2$ cannot coalesce 
without equaling $z$.
The corresponding density $f_1$ is also obtained as a superposition: setting
$$
w(j,k)=g(\eta_k+\zeta)\,(1+\theta(\eta_k+\zeta))\,\exp
i(\eta_k-\zeta)x+\eta_{j}y)
$$
we find
$$
f_1(x,y)=
\cases 
w(1,2)-w(2,1), & y<x ;\\
-\theta(\eta_{2}+\zeta)w(2,1)+\theta(\eta_{1}+\zeta)w(1,2),
& x<y.
\endcases
$$

\proclaim{Theorem 5.1} The antisymmetrized wave functions (5.1) and (5.2) 
satisfy the regularity condition (2.5) and thus determine a basis of
eigenstates for the Hamiltonian (2.1).
\endproclaim
\demo{Proof} Consider a pair $(y_k,y_{k+1})$ (the proof is the same for 
contiguous $z$'s) and assume there are no
intervening $x's$ or $z's$. Group the terms in the sum (5.1) into pairs 
corresponding
to $J=(\dots,j,j+1\dots)$ and $J'=(\dots j+1,j\dots)$. At $y_j=y_{j+1}$
we have $e_0(J,K)=e_0(J',K)$, and the coefficient of this common value is
$$
\phi^0_{\s}(J,K)-\phi^0_{\s}(J',K).
$$
 
For any given pair $(j,k)$ either $y_j<y_{j+1}<z_k$ or
$z_k<y_j<y_{j+1}$. In the first case, neither $(j,k)$ nor $(j+1,k)$
belong to
$T_{\s}$, so neither of the factors $\theta (\eta_{J_j}-\zeta_{K_k})$,
$\theta(\eta_{J_{j+1}}-\zeta_{K_k})$ belong to the products 
$\phi^0_{\s}(J,K)$, $\phi^0_{\s}(J',K)$.
In the second case, both $(j,k)$ and $(j+1,k)$ belong to $T_{\s}$, so
both factors belong to both the products, and $\phi^0_{\s}(J,K)=
\phi^0_{\s}(J',K)$. Therefore the coefficient of $e_0(J,K)=e_0(J',K)$ 
vanishes in either case, and
the sum is zero. This establishes the result for $f_0$.

We proceed in the same way to verify that the higher 
order densities $f_m,\ m\ge 1$ vanish
when $y_j$ and $y_{j+1}$ coincide. In view of (4.16), we must
show that
$$
\phi^m_{\s}(J,K)g_{\s}(J,K)=\phi^m_{\s}(J',K)g_{\s}(J',K).
$$
{From} (4.12) the constants $g_{\s}(J,K)$ depend only on the momenta 
associated with the bosons (the zeroes of $\s$), and so are unchanged under 
any permutation of the fermion momenta. The same applies to the product in 
(4.13): it depends only on the boson, not the fermion momenta. Therefore the
issue for the higher order densities reduces to the proposition
$\phi^0_{\Sigma}(J,K)=\phi^0_{\Sigma}(J',K)$, which
was established in the first step of this proof. \qed
\enddemo

Given the $f_m$ defined on the cells 
$$
\{\ 0\le x_1<\dots <x_m\le L, \ 0\le y_1<\dots<y_n\le L , \ 0\le \ 
z_1<\dots<z_p\le L\ \},
$$
we extend them to the hypercube $0\le x_i,\,y_j,\,z_k\le L$ by
$$
h_m(x;y;z)=\text{sgn}(J)\text{sgn}(K)f_m(Mx,Jy,Kz) \tag5.3
$$
where $Mx,\ Jy,\ Kz$ are the unique permutations of the variables $x,y,z$ 
for which
$$
x_{M_1}<\dots <x_{M_m},
\qquad
y_{J_1}<\dots<y_{J_n},
\qquad
z_{K_1}<\dots<z_{K_p}.
$$

We now impose periodic boundary conditions on the wave functions $h_m$. 
These periodicity conditions lead to constraints on the momenta, known as 
the Bethe equations. 

For example, in the simplest case $N=2,\ D=0$ (cf. \S4) we obtained
$$
f_0(y,z)=\cases \exp\, i(\eta y-\zeta z), & y<z;\\
\theta(\eta+\zeta)\,\exp\, i(\eta y-\zeta z), & y>z.
\endcases
$$
The periodicity conditions 
$$
f_0(0,z)=f_0(L,z),
\qquad
f_0(y,0)=f_0(y,L)
$$
lead immediately to the Bethe equations
$$
1=\theta(\eta+\zeta)e^{i\eta L}
\qquad
e^{i\zeta L}=\theta(\eta+\zeta).
$$

For the case $n=2,\ p=1$, the conditions (5.3) lead to
$$
f_0(y_1,y_2,0)=f_0(y_1,y_2,L)
\qquad
f_0(0,y_2,z)=-f_0(y_2,L,z).
$$
{From} the definition of $f_0$ in this case we get, respectively,
$$
\align
\theta(\eta_1+\zeta)\,\theta & (\eta_2+\zeta)\,(\exp\,i(\eta_1y_1+\eta_2y_2)
-\exp\,i(\eta_1y_2+\eta_2y_1) )\\
&=\exp\, i(\eta_1y_1+\eta_2y_2-L\zeta)-\exp\, i(\eta_1y_2+\eta_2y_1-L\zeta)
\endalign
$$
and (say for $y_2<z$)
$$
\align
\exp\,i(\eta_2y_2-\zeta z)-\exp\,i(\eta_1y_2-\zeta z)
&= -\theta (\eta_2+\zeta)\,\exp\,i(\eta_1y_2+\eta_2L-\zeta z)\\
&\quad+\theta(\eta_1+\zeta)\,\exp\,i(\eta_2y_2+\eta_1L-\zeta z).
\endalign
$$
{From} these two equations we get the Bethe equations
$$
\gather
\exp\, iL\zeta=\theta(\eta_1+\zeta)\,\theta(\eta_2+\zeta);\\
1=\theta(\eta_1+\zeta)\,\exp\, i\eta_1L,
\qquad
1=\theta(\eta_2+\zeta)\,\exp\, i\eta_2L.
\endgather
$$
All other periodicity conditions follow {from} these.

\demo{Remark} The second example shows that one 
must anti-symmetrize in momenta to achieve periodicity. 
For example, suppose
$N=3$, $D=1$, and $0<y_2<z$.  The condition $h_0(0,y_2,z)=-h_0(y_2,L,z)$ 
for the basic wave function obtained in \S4 becomes
$$
h_0(0,y_2,z)=\exp\,i(\eta_2y_2-\zeta z)=-
h_0(y_2,L,z)=\theta(\eta_2+\zeta)\,\exp\,i(\eta_1y_2+\eta_2L-\zeta z),
$$ 
leading to the contradiction
$$
\exp\,i\eta_2y_2=-\theta(\eta_2+\zeta)\,\exp\,i(\eta_1 y_2+\eta_2 L).
$$\enddemo

\proclaim{Theorem 5.2} Let
$$
\align
e^{i\eta_k L}=&\prod_{l=1}^p \theta (\zeta_l+\eta_k) 
\qquad  k=1, \dots,n, \tag5.4\\
e^{i\zeta_l L}=&\prod_{k=1}^n \theta(\eta_k+\zeta_l) 
\qquad
l=1, \dots,p \tag5.5
\endalign
$$
Then the densities $h_m$ defined by (5.1), (5.2), and (5.3) are
 $L$-periodic in each variable.
\endproclaim

The equations (5.4), (5.5) are known as the Bethe equations. 
The proof is given in Beals, Sattinger, and Williams$^{15}$.

\ms\ni
{\bf 6.  CONSTRUCTION OF A DIRAC GROUND STATE} 
\ss
A solution of the Bethe equations gives an eigenvector of the Hamiltonian
with energy 
$$
E=-c_1\sum_{k=1}^n\eta_k+c_2\sum_{l=1}^p\zeta_l,\tag 6.1
$$
where $n$ and $p$ denote the maximum number of particles of types 1 and 2 for 
states in the 
space $\Cal F_{N,D}$, i.e. $n=\tfrac12(N+D)$, $p=\tfrac12(N-D)$; cf. (4.5).
The energy is not bounded below.
We construct a stable ground state by
filling all  negative energy states.  

Taking logarithms in the Bethe equations (5.4), (5.5), we get
$$
\gather
\eta_k -\frac{2\pi}{L} I_k = \sum_{l=1}^p \frac{1}{iL}\log \theta(\eta_k
+\zeta_l), 
\qquad k=1,\dots,n\tag 6.2\\ 
\zeta_l -\frac{2\pi}{L} J_l = \sum_{k=1}^n \frac{1}{iL}\log \theta(\eta_k
+\zeta_l)\qquad l=1,\dots,p
\endgather
$$
where
$$
\theta(\l)=\frac{\l+i\omega}{\l-i\omega},
\qquad
\omega=\frac{\g^2}{4\alpha^2}\tag 6.3
$$
and $I_k$ are distinct integers, as are $J_l$.

Taking
$$
\frac{1}{i}\log \theta(\l)=\pi -2 \tan^{-1}\frac{\l}{\omega},\tag 6.4
$$
we write the Bethe equations in the equivalent form
$$
\gather
\eta_k =\frac{2\pi}{L}I_k,+
\frac{2}{L} \sum_{l=1}^p \tnin \frac{\eta_l+\zeta_l}{\omega},\tag 6.5\\
\zeta_l=\frac{2\pi}{L}J_l+ \frac{2}{L} \sum_{k=1}^n
\tnin \frac{\eta_k+\zeta_l}{\omega}
\endgather
$$
where the $I_k$ are distinct half integers if $p$ is odd, 
distinct integers if $p$ is even, and similarly for the $J_l$
in relation to $n$.   We assume for convenience that
both are even and, without loss of generality, that 
$$
I_1<I_2<\dots<I_n;\quad J_1<J_2<\dots<J_p.\tag 6.6
$$
These conditions imply that there is a solution to (6.5) for which the
$\eta_k$ and the $\zeta_l$ are themselves strictly increasing.
Indeed equations (6.7) are the variational equations for the functional
$$
\Sigma=\frac12 \left(\sum_{k=1}^n \eta_k^2+\sum_{l=1}^p \zeta_l^2 \right)+
\frac{\pi}{L}\left(\sum_{k=1}^n I_k\,\eta_k +\sum_{l=1}^p J_l\,\zeta_l \right)
-\sum_{k,l=1}^{n,p}M(\eta_k+\zeta_l)
$$
where 
$$
M(\l)=\int_0^{\l}\frac{2}{L}\tan^{-1}\frac{s}{\omega}\,ds.
$$

For convenience in this section we take $\omega=1$.  (This amounts to fixing 
a length scale, since $\omega$ has the dimension 1/length.)
\ss
\proclaim{Theorem 6.1} The Bethe equations have at least one solution
for each set of distinct integers $I_k$ and distinct integers $J_l$.
\endproclaim
\ss
\demo{Proof}  The term $M(\l)$ grows only linearly in its argument, 
so the quadratic terms dominate $\Sigma$ as $\eta,\,\zeta \tti$. It follows 
that $S$ is bounded below and tends to infinity for large values of its 
argument.  It therefore always has a minimum.\quad \qed
\enddemo
\ss
Uniqueness of the solutions is a subtler question. For large values of 
the densities,  $\Sigma$  is not convex, so in principle it could have 
multiple 
critical points.  In the limiting cases of interest, however, we will obtain
uniqueness results.  In fact we use the finite case to motivate the limiting 
cases. Therefore we save space and smooth the exposition by 
stating most of the finite case results without proof and leaving mathematical 
details for the limiting cases.
\ss
We now turn to the construction of a stable ground state, or vacuum state. 
For the ground state we assume  that $D=0$ (hence $n=p=N/2$) and we consider
the case when the velocities of the fermions have opposite signs:
$$
c_2<0<c_1.\tag 6.7
$$
Thus $\eta>0$ and $\zeta>0$ mean negative energy for the corresponding
particles.  {\it Our procedure will be to fill the available negative energy
states {from} the top down\/} by choosing consecutive integers $I_k$, $J_l$
such that the solution $\{\eta_k,\z_l\}$ is positive {\it but minimally so\/};
we then take the infinite length and infinite density limits.  The
equations (6.5) are symmetric, so we take the data $I_k=J_k$ and look for
a solution with $\eta_k=\z_k$.  Thus we want
$$
I_k=J_k=k-\kappa_n\tag 6.8
$$
where $\kappa_n$ is an integer to be chosen.

\proclaim{Theorem 6.2} There is a largest integer $\k_n=O(n)$ such that
(6.5) with data (6.8) has a positive solution.  This solution is unique and 
satisfies $\eta_k=\z_k$, and $\eta_1<\dots<\eta_n$.  Moreover, $\eta_1\to 0$
as $n\to\infty$, $L\to\infty$  with $n/L$ bounded above. 
\endproclaim

We refer to the solution in Theorem 6.2 as the {\it cut-off
ground state\/} (CGS).  Our interest is the infinite length, finite
density limit, and then the infinite density limit. For this purpose 
we recast (6.5) as a pair of integral equations.  
Assuming still that $n=p$, we let $\Delta=2\pi n/L$ and
define functions on the interval $[0,\Delta)$ by
$$\gather
I(\xi)=\tfrac{2\pi}L I_k,\quad J(\xi)=\tfrac{2\pi}L J_k,\quad
\tfrac{2\pi(k-1)}L\le\xi<\tfrac{2\pi k}L;\tag 6.9\\
\eta(\xi)=\eta_k,\quad \zeta_L(\xi)=\zeta_k\quad
\tfrac{2\pi(k-1)}L\le\xi<\tfrac{2\pi k}L.
\endgather
$$
(We hope that context will prevent confusion
between the use of $\eta$ and $\z$ to denote {\it functions\/} and
the use of $\eta$ as $\z$ as pure variables.)

Equations (6.2) with $n=p$ are equivalent to integral equations 
$$
\align
\eta(\xi)&=I(\xi)+\frac1\pi\int^{2\pi n/L}_0 
\tnin(\eta(\xi)+\zeta(\xi'))\,d\xi,\tag 6.10\\
\zeta(\xi)&=J(\xi)+\frac1\pi\int^{2\pi n/L}_0
\tnin(\eta(\xi')+\zeta(\xi))\,d\xi'.
\endalign
$$
It can be shown that for every set of data $I,J$ corresponding to
sets of integers indexed as in (6.6), there is a solution $(\eta,\z)$
with positive jumps at each of the points of discontinuity $2\pi k/L$.
We define approximate inverse functions accordingly:
$$
\xi(\eta)=\inf\{\xi:\eta(\xi)\ge \eta\};\quad
\xi(\z)=\inf\{\xi:\z(\xi)\ge\z\}.\tag 6.11
$$
\demo{Remark} These functions have jumps of size $2\pi/L$, so their difference
quotients with respect to $\eta$ and to $\z$ respectively count
the number of particles associated to a given momentum interval.\enddemo

With the Bethe equations in the form (6.10), it is clear how to pass to the 
infinite length, finite
density limit.  To pass also to the infinite density limit we observe
that (6.10) is equivalent to
$$
\align
\eta(\xi)-\eta(0)&=I(\xi)-I(0)\tag 6.12\\
&+\frac1\pi\int^{2\pi n/L}_0[\tnin(\eta(\xi)+\zeta(\xi'))
- \tnin(\eta(0)+\zeta(\xi'))]\,d\xi';\\
\zeta(\xi)-\z(0)&=J(\xi)-J(0)\\
&+\frac1\pi\int^{2\pi n/L}_0[\tnin(\eta(\xi')+\zeta(\xi))
-\tnin(\eta(\xi')+\zeta(0))]\,d\xi'.
\endalign
$$
The data (6.8) give rise to the functions $I(\xi)=J(\xi)=\xi-c_\D$ in the
$L\to\infty$ limit, $\D$ finite, and in this limit $\eta(0)=z(0)=0$. Thus
in the form (6.12) we 
pass to the $\D=\infty$ limit and find the equations
for the {\it ground state\/} $(\eta_0, \z_0=\eta_0)$:
$$
\eta_0(\xi)=\xi
+\frac1\pi\int^\infty_0[\tnin(\eta_0(\xi)+\eta_0(\xi'))
- \tnin(\eta_0(\xi'))]\,d\xi'.\tag 6.13
$$

We obtain existence and uniqueness of the solution to (6.13) by converting
it to a linear integral equation for the derivative of the inverse 
function:
$$
\r_0(\eta(\xi))=\frac1{(d\eta_0/d\xi)(\xi)}.\tag 6.14
$$
Differentiate (6.9), multiply by $\r_0$, and change the variable
of integration to obtain:
$$
\rho_0(\eta)=1-\frac1{\pi}\int_0^\infty\frac{\rho_0(\eta')\,d\eta'}
{1+(\eta+\eta')^2}=1-K(\r_0).\tag 6.15
$$

\demo{Remarks} 1. In view of the remark after (6.9),  $\r_0$ may be 
thought of as the {\it particle density function\/} for
the ground state.  

2. Note that the kernel of the integral operator $K$ satisfies
$$
\sup_{\eta\ge 0}\ \frac{1}{\pi}\int_0^\infty\frac{d\eta'}{1+(\eta+\eta')^2}
 = \frac12.\tag 6.16
$$
Therefore $T$ has norm $1/2$ as an operator in  $L^{\infty}(\R_+)$
and (6.16) has a unique solution, obtainable by successive
approximations.  The solution is positive and bounded.  \enddemo

In the next section we need both the analytic continuation $\r_{0,t}$ of $\r$
to the negative axis and a dual function $\rs$ that is the analytic
continuation of the solution to the same equation but with a change of sign:
$$
\rho_0^*(\eta)=1+\frac{1}{\pi}\int^\infty_0 \frac{\rho^*_0(\eta')\,d\eta'}
{1+(\eta+\eta')^2}=1+K\rs(\eta),\eta\ge 0.\tag 6.17
$$
In terms of these functions we define functions $\xi_0$ and
$\xi^*_0$ 
$$
\xi_0(\eta)=\int^\eta_0\r_{0,t}(s)\,ds;\quad \xi^*_0(\eta)=\int_0^\eta
\rs(s)\,ds,\quad\eta\in\R.\tag 6.18
$$

\proclaim{Theorem 6.3} The extension $\r_{0,t}$ of the particle density 
function $\r_0$ and the dual function $\rs$ satisfy the estimates
$$\gather
0<\rho_{0,t}(\eta)<1<\rs<2,\quad \eta\in\R;\tag 6.19\\
\r_{0,t}(\eta)=1+O(1/\eta), \quad \rs(\eta)=2+O(1/\eta)\quad\text{as}\quad
\eta\to\infty;\\
\pi\eta\,\r_{0,t}(\eta)=-1+O(\log|\eta|/|\eta|),\quad
\rs=1+O(1/\eta) \quad\text{as}\quad \eta\to-\infty.
\endgather
$$
The functions $\xi_0$ and $\xi^*_0$ have asymptotics
$$\gather
\xi_0(\eta)=\eta+O(1),\quad\xi^*_0(\eta)=2\eta+O(1)
\quad\text{as}\quad\eta\to+\infty;\tag 6.20\\
\pi\,\xi_0(\eta)=-1+O(\log|\eta|/|\eta|),\quad\xi^*_0(\eta)=\eta+O(1)
\quad\text{as}\quad\eta\to-\infty.\endgather
$$
In particular the function $\xi_0$ is a diffeomorphism of the line.
Its inverse function $\eta_0$ satisfies the integral equation (6.13)
on $\R_+$.
\endproclaim
\demo{Proof} It follows {from} (6.16) that the solution  
extends analytically.  The kernel is positive, so
$0<\rho_0<1$.  The asymptotics of $\rho_0$ and of $\xi_0$ are
easily verified {from} (6.17).  The argument for $\rs$ and $\xi^*_0$
are similar.
Finally, (6.13) results {from} substituting (6.16) in the definition
of $\xi_0$ and
using $\r_0\, d\eta'=d\xi'$ to change the variable of integration.
\quad\qed\enddemo

We conclude this section with the integral expressions
of various quantities associated with a
state described by a pair of functions $\eta,\z$.  
For a finite number of particles the momentum $P$ and
energy $E$ are 
$$\gather
P=\sum^n_{k=1}\eta_k-\sum^n_{l=1}\z_l=\frac{L}{2\pi}\int^{2\pi n/L}_0
\eta(\xi)\,d\xi-\frac{L}{2\pi}\int^{2\pi n/L}_0\z(\xi)\,d\xi;\tag 6.21\\
E=-c_1\sum^n_{k=1}\eta+c_2\sum^n_{l=1}\z_l=
-\frac{c_1L}{2\pi}\int^{2\pi n/L}_0
\eta(\xi)\,d\xi+\frac{c_2L}{2\pi}\int^{2\pi n/L}_0\z(\xi)\,d\xi.\endgather
$$
The particle number $n=p$ itself can be expressed in terms
of the inverse functions: 
$$
n=\frac{L}{2\pi}\sup\xi(\eta)=\frac{L}{2\pi}\sup\xi(\z).\tag 6.22
$$
Therefore in the infinite length limit the particle densities per unit length
are 
$$
\frac1{2\pi}\sup\xi(\eta)=\frac1{2\pi}\int \r(\eta)\,d\eta;\quad
\frac1{2\pi}\sup\xi(\zeta)=\frac1{2\pi}\int \s(\zeta)\,d\zeta.\tag 6.23
$$

\ss\ni
{\bf 7. FINITE PARTICLE EXCITED STATES}

\ss
In this section we construct particle-hole excitations of the ground
state.  These have finite positive energy with respect to
the ground state in the infinite density limit, confirming the
stability of the ground state.  

We begin with a single particle/hole excited state
for an $\eta$ particle, with momentum $\eta_p<0$ for the particle and
$\eta_h>0$ for the hole.  Consider first a cut-off ground state
with a high particle density, so that there are large momentum particles.
Denote the functions for this CGS by $\eta,\z=\eta$.
We change the CGS data by choosing $k$ such that $\eta_k$ is close to
$\eta_h$ and changing $I_k$ to $I_k-s$, where the integer $s$ is
chosen so that $I_k-s$ is close to $L\xi_0(\eta_p)/2\pi$.  In the integral
equation picture (6.10) the new data is given by functions $\bar I$,
$\bar J=I$.  Here $\bar I=I$ except on an interval of length $2\pi/L$
located near $\xi_h=\xi_0(\eta_h)$, on which it differs {from} $I$ by
approximately $\xi_p-\xi_h$, $\xi_p=\xi_0(\eta_p)$. This change in data
changes the $\eta$ solution by a bounded amount on that interval and
otherwise changes both $\eta(\xi)$ and $\z(\xi)$ by $O(1/L)$.  Therefore
we write the solution $(\bar\eta,\bar\z)$ of the perturbed equation as
$$
\bar\eta(\xi)=\eta(\xi)+e_1(\xi)/L+f(\xi),\quad \bar\z(\xi)=
\z(\xi)+e_2(\xi)/L\tag 7.1
$$
where $e_1$ vanishes on the critical interval and $f$ is supported on
the critical interval.  With a similar notation for energy and momentum,
the excitation momentum and energy are
$$\align
\delta P&=\bar P-P=\frac1{2\pi}\int^\D_0[e_1(\xi)-e_2(\xi)]\,d\xi+f;
\tag 7.2\\
\delta E&=\bar E-E=\frac1{2\pi}\int^\D_0[c_1e_1(\xi)-c_2e_2(\xi)]\,d\xi
+c_1f.\endalign
$$
Here we abuse notation and write $f$ also for the {\it value\/}
of $\bar\eta-\eta$ on the critical interval.

With large $L$ the integral equations (6.10)
give 
$$\align
e_1(\xi)&\approx \frac1\pi\int^\D_0\frac{[e_1(\xi)+e_2(\xi')]\,d\xi'}
{1+(\eta(\xi)+\eta(\xi'))^2};\tag 7.3a\\
f&\approx \xi_p-\xi_h +\frac1\pi\int^\D_0[\tnin(\eta_h+f+\eta(\xi'))-
\tnin(\eta_h+\eta(\xi'))]\,d\xi';\tag 7.3b\\
e_2(\xi)&\approx 2[\tnin(\eta_h+f+\eta(\xi))-\tnin(\eta_h+\eta(\xi))]\\
&\qquad+\frac1\pi\int^\D_0\frac{[e_1(\xi')+e_2(\xi)]\,d\xi'}
{1+(\eta(\xi)+\eta(\xi'))^2}.\tag 7.3c\endalign
$$
Here (7.3a) is valid outside the critical interval.
 
In the infinite length, infinite density limit (6.13) and (7.3b) show
that $f=\eta_h-\eta_p$, so $\bar\eta\approx\eta_h$ on the critical
interval, as desired.  In this limit the remaining equations (7.3)
become 
$$\align
e_1(\eta)&=\frac1\pi\int^\infty_0\frac{[e_1(\eta)+e_2(\eta')]\,\r_0(\eta')
\,d\eta'}{1+(\eta+\eta')^2};\tag 7.4\\
e_2(\eta)&=\frac1\pi\int^\infty_0\frac{[e_1(\eta')+e_2(\eta)]\,\r_0(\eta')
\,d\eta'}{1+(\eta+\eta')^2}\\
&\qquad+2\tnin(\eta_p+\eta)-2\tnin(\eta_h+\eta).\endalign
$$
We write $\tilde e_j(\eta)$ for $e_j(\xi_0(\eta))\r_0(\eta)$:
The integral equation $(6.18)$ for $\r_0$ allows (7.4) to be rewritten:
$$\align
\tilde e_1(\eta)&=K(\tilde e_2)(\eta);\tag 7.5\\
\tilde e_2(\eta)&=2\tnin(\eta_p+\eta)-2\tnin(\eta_h+\eta)+T(\tilde e_1)(\eta).
\endalign
$$

In the limit the excitation momentum and energy are
$$\align
\delta P &= \frac1{2\pi}\int_0^\infty [\tilde e_1(\eta)-\tilde e_2(\eta)]\,
\,d\eta+(\eta_p-\eta_h);\tag 7.6\\
\delta E& = \frac1{2\pi}\int^\infty_0[-c_1\tilde e_1(\eta)
+c_2\tilde e_2(\eta)]\,d\eta-c_1(\eta_p-\eta_h)\\
&=-{\alpha}\left(\frac1{2\pi}\int^\infty_0
[\tilde e_1(\eta)+\tilde e_2(\eta)]\,d\eta+
\eta_p-\eta_h\right)\\
&\qquad\qquad+{c_0}\left(\frac1{2\pi}\int^\infty_0
[\tilde \e_1(\eta)-\tilde e_2(\eta)]\,d\eta-\eta_p+\eta_h\right).
\endalign
$$
(The last part of (7.6) follows {from} the second part, since $2\alpha=
c_1-c_2$ and $2c_0=c_1+c_2$.)

The system (7.5) and the equations (7.6) thus characterize this single
particle/hole excitation of the ground state.  Again the fact that
$K$ has norm $1/2$ implies that (7.6) has,
for each choice of $\eta_h$ and $\eta_p$, a unique solution.  Positivity
of the kernel implies that the solution $(e_1,e_2)$ is negative when
$\eta_p-\eta_h$ is negative, so the excitation energy is positive.
We can be more precise.

\proclaim{Theorem 7.1} 
The excited state momentum and the excitation 
energy can be expressed in terms of the (extended) functions $\xi_0$,
$\xi^*_0$ of Theorem 6.3.  They have the form
$$\align
\delta P &=  P_p- P_h=  P(\eta_p)- P(\eta_h)
=\xi_0(\eta_p)-\xi_0(\eta_h);\tag 7.7\\
\delta E &=E_p-E_h=E(\eta_p)-E(\eta_h)\\
&=[-\alpha\xi^*_0(\eta_p)-c_0\xi_0(\eta_p)]
-[-\alpha\xi^*_0(\eta_h)-c_0\xi_0(\eta_h)].\endalign
$$
These functions have the asymptotics
$$\align
E_p &\sim 2\alpha e^{\pi|P_p|}\quad\text{as}\quad P_p\to-\infty,\tag 7.8\\
E_h &\sim -c_1  P_h\quad\text{as}\quad P_h\to+\infty.\endalign
$$
\endproclaim
\demo{Proof} The inhomogeneous
term for the pair $(\eta_p,\eta_h)$ is the sum
of the inhomogeneous term for the pair $(\eta_p,0)$ and the inhomogeneous
term for the pair $(0,\eta_h)$ and is an odd function of its two
arguments.  Therefore it suffices to consider the case $\eta_p=\te$, 
$\eta_h=0$.   We make the ${\te}$ dependence explicit by writing
$\tilde e_1(\eta,\te)$, $\tilde e_2(\eta,\te)$.  Set
$$
f_1=\frac1{2\pi}\pd{\te}\,[\tilde e_1+\tilde e_2];\qquad f_2=
\frac1{2\pi}\pd{\te}\,[\tilde e_2-\tilde e_1].\tag 7.9
$$
(Note that the $\tilde e_j$ vanish at $\tilde\eta=0$, so the sum and difference
can be recovered by integrating the $f_j$ {from} $\tilde\eta=0$.)
Adding and subtracting first derivatives with respect to $\te$ in
(7.5), we find that as functions of $\eta$ the $f_j$ satisfy
$$
f_1=Kf_1+k,\qquad f_2=-Kf_2+k,\qquad k(\eta,\te)=\frac1\pi\,\frac1
{1+(\eta+\te)^2}.\tag 7.10
$$
Note that $k$ is the kernel of the operator $K$.  Let $F_j$ denote 
the integral operator that has kernel $f_j$ and $I_d$ the identity
operator.  Then (7.10) is equivalent to
the operator equations
$$
(Id-K)(Id+F_1)=Id,\qquad (Id+K)(Id-F_2)=Id.\tag 7.11
$$
The integral $\int^\infty_0 f_j(\eta,\te)\,d\eta$ is the same 
as the function $F_j(1)(\te)$,
where $1$ denotes the function that is identically $1$.  It follows 
{from} (7.9) and (7.10) that
$$\align
\frac1{2\pi}\int^\infty_0 \tilde e_1&(\eta,\te)\,d\eta+\te=
\frac12\int^{\te}_0\bigg[\int^\infty_0(f_1-f_2)+2\bigg]\,d\eta\tag 7.12\\
&=\frac12\int^{\te}_0 [(F_1 (1)+ 1)+(1 -F_2 (1))];\\
\frac1{2\pi}\int^\infty_0 \tilde e_2&(\eta,\te)\,d\eta=
\frac12\int^{\te}_0\bigg[\int^\infty_0(f_1+f_2)\bigg]\,d\eta\\
&=\frac12\int^{\te}_0 [F_1 (1)  + F_2 (1)]\,d\eta
=\frac12\int^{\te}_0 [(F_1(1)+1)  -(1- F_2(1))]\,d\eta
\endalign
$$
In view of (7.11), $F_1(1)+1$ is the solution of $u=Ku+1$, while
$1-F_2(1)$ is the solution of $v=-Kv+1$.  By (6.17) and (6.16) these
functions are $\rs$ and $\r_0$, respectively.  It follows that
$$\align
\frac1{2\pi}\int^\infty_0 \tilde e_1(\eta,\te)&\,d\eta+\te=
\tfrac12[\xi^*_0(\te)+\xi_0(\te)],\tag 7.13\\
\frac1{2\pi}\int^\infty_0 \tilde e_2(\eta,\te)&\,d\eta=
\tfrac12[\xi^*_0(\te)-\xi_0(\te)].\tag 7.14\endalign
$$
The momentum term $P(\te)$ in (7.6) is the difference of the two
expressions (7.13) and (7.14).
Similarly, the excitation energy term $E(\te)$ in (7.6) is
is the linear combination of (7.13) and (7.14)
with coefficients $-c_1$ and $c_2$ respectively.  This
proves (7.7).  The asymptotics (7.8) follow {from} (7.7) and
the asymptotics (6.20) of $\xi_0$ and $\xi_0^*$.\quad\qed\enddemo

Exactly similar considerations apply to a single particle/hole $\z$
excitation, with the roles of $e_1$ and $e_2$ interchanged.  In particular,
the indices $1$ and $2$ in (7.9) should be interchanged throughout.
\smallskip

Compound excitations may be treated in exactly the same way.  Consider the case
of particle/hole excitations in $\eta$ described by
$\eta_{h,i}\ge 0$, $\eta_{p,i}\le 0$,
$i=1,\dots,r$ and in $\zeta$ described by $\zeta_{h,j}
\ge 0$, $\zeta_{p,j}\le 0$, $j=1,\dots,s$.
Then in the infinite density limit the Bethe equations become 
$$\align
\tilde e_1(\eta)&=K(\tilde e_2)(\eta)
+2\sum^s_{j=1}[\tnin(\eta+\zeta_{p,j})
-\tnin(\eta+\zeta_{h,j})],\tag 7.15\\
\tilde e_2(\eta)&=K(\tilde e_1)(\eta)
+ 2\sum^r_{i=1}[\tnin(\eta+\eta_{p,i})-\tnin(\eta+
\eta_{h,i})].\endalign
$$
This state is a superposition of corresponding single particle/hole
excited states.  Therefore
the associated momentum and excitation energy are given by
$$\align
\delta P&=\sum^r_{i=1}[\xi_0(\eta_{p,i})-\xi_0(\eta_{h.i})]+
\sum^s_{j=1}[\xi_0(\zeta_{h,j})-\xi_0(\zeta_{p,j})];\tag 7.16\\
\delta E&=-\alpha
\bigg(\sum^r_{i=1}[\xi_0^*(\eta_{p,i})-\xi_0^*(\eta_{h,i})]+\sum^s_{j=1}
[\xi_0^*(\zeta_{p,j})-\xi_0^*(\zeta_{h,j})]\bigg)\\
&\qquad +c_0\bigg(\sum^r_{i=1}[\xi_0(\eta_{p,i})-\xi_0(\eta_{h,i})]
-\sum^s_{j=1}[\xi_0(\zeta_{p,j})-\xi_0(\zeta_{h,j})]\bigg).\endalign
$$
\ss\ni
{\bf 8. FINITE DENSITY PERTURBATIONS OF THE GROUND STATE}
\ss
In this section we study large 
deviations {from} the ground state.  Again we motivate
the procedure by looking briefly at the finite length, finite particle
case in the form of the integral equations (6.10).  Denote by $I_0=J_0$
the data that corresponds to the CGS data (6.8).  The CGS data
uses all the integers {from} an interval with endpoints $\pm O(n/2)$
and we are temporarily applying a low energy cut-off that corresponds
to the upper end of this interval, so the data that corresponds
to an admissible perturbation of the CGS satisfies
$I,J\le I_0$.  The solution can be found by iteration from the
integral equations (6.10).  The monotonicity properties of the
mapping imply that the solution is dominated by the CGS $(\eta_0,\z_0)$:
$$
\eta(\xi)\le\eta_0(\xi),\quad\z(\xi)\le\z_0(\xi). \tag 8.1
$$
Moreover if the perturbation is to have finite relative energy
density in the limit,
we expect that $I,J\approx I_0$ for large $\xi$.  Note also that
the jumps in the data must be at least as great as
the jumps $2\pi/L$ in $I_0$.   We pass to the infinite length,
infinite (total) density limit and look for a solution of an
analogue of (6.12):
$$\align
\eta(\xi)-I(\xi)
+\frac1\pi\int^\infty_0[\tnin(\eta(\xi)+\zeta(\xi'))
- \tnin(\eta_0(0)+\zeta_0(\xi'))]\,d\xi';\tag 8.2\\
\zeta(\xi)-J(\xi)
+\frac1\pi\int^\infty_0[\tnin(\eta(\xi')+\zeta(\xi))
- \tnin(\eta_0(\xi')+\zeta_0(\xi))]\,d\xi'.\endalign
$$
According to the preceding discussion the natural assumptions 
on the data are
$$
\frac{dI}{d\xi}\ge 1,\quad\frac{dJ}{d\xi}\ge 1;\quad
I(\xi)\approx J(\xi)\approx \xi,\ \ \ \xi>>0.\tag 8.3
$$
It can be shown that finite particle solutions whose data is such that 
the formal limits are (8.2) converge to solutions of (8.2).  However it
is more convenient for what follows to write the equations for the
particle density functions
$\r =d\xi/d\eta$, $\s=d\xi/d\z$:
$$\align
\r(\eta)\cdot I'(\eta)&=1-\frac1\pi\int^\infty_{-\infty}\frac{\s(\z)\,d\z}
{1+(\eta+\z)^2}=1-(K\s)(\eta);\tag 8.4\\
\s(\z)\cdot J'(\z)&=1-\frac1\pi\int^\infty_{-\infty}\frac{\r(\eta)\,d\eta}
{1+(\eta+\z)^2}=1-(K\r)(\z),\endalign
$$
where 
$$
I'(\eta)=\frac{dI}{d\xi}(\xi(\eta));\quad J'(\z)=\frac{dJ}{d\xi}(\xi(\z)).
$$
We have removed the upper bound on particle energies, so the integrations
in (8.4) need to be taken over the whole line.  Now
$K$ has norm 1 as an operator in $L^\infty(\R)$ so solvability of (8.4) 
becomes an issue.  We rewrite (8.4) in the form
$$
\r=\t(1-K\s),\quad \s=\Psi(1-K\r)\tag 8.4'     
$$
and adopt the point of view that the functions $\t$ and $\Psi$ are the
data.  For the ground state the data are $\Psi_0=\t_0$, the Heaviside
function: $\t_0(s)=0$, $s<0$; $=1$, $s\ge0$.  
More generally we can express the conditions (8.3) and
the assumption that our states are perturbations of the ground state
by requiring
$$
0\le \t,\ \Psi\le 1;\quad \lim_{|\eta|\to\infty}(\t-\t_0)=\lim_{|\z|\to\infty}
(\Psi-\Psi_0)=0,\tag 8.5
$$
while (8.1) gives an implicit constraint on the data $(\t,\Psi)$ in 
terms of the solution.  In fact (8.1) implies the reverse relation
between the inverse functions, leading to 
$$
\int_{-\infty}^\eta (\r-\r_0)\ge 0,\quad\int_{-\infty}^\z (\s-\s_0)\ge
0, \quad\text{all } \eta,\s.\tag 8.6
$$
Moreover the cut-off solution $(\eta,\z)$ and the corresponding
CGS $(\eta_0,\z_0)$ at any given step were defined over the same
$\xi$ interval, so the relative particle density should be zero:
$$
\int_{-\infty}^\infty (\r-\r_0)= 0=\int_{-\infty}^\infty
(\s-\s_0).\tag 8.7
$$
We show below that (8.7) is necessary for the relative energy density
to be finite.

\proclaim{Theorem 8.1} Under assumption (8.5) the equations (8.4)
have a unique solution.  For the solution we have
$$
0\le \r\le 1,\quad 0\le\s\le 1;\tag 8.8
$$
$$
\lim_{\eta\to\infty}\r(\eta)=\lim_{\z\to\infty}\s(\z)=1;\quad
\lim_{\eta\to-\infty}\r(\eta)=\lim_{\z\to-\infty}\s(\z)=0.
$$
\endproclaim

\demo{Proof} We use $\t$ and $\Psi$ also to denote the operations 
of multiplication by the corresponding functions.  Then the equations
are
$$
\r+\t K\s=\t,\quad\s+\Psi K\r=\Psi.
$$
We apply the operator $Id-\Psi K$ to the first equation and 
$Id-\t K$ to the second and reorganize to obtain
$$
(Id-\t K\Psi K)\r=\t-\t K\Psi,\quad (Id-\Psi K\t K)\s=\Psi-\Psi K\t .\tag 8.9
$$
It is enough to show that $\t K\Psi$ and $\Psi K\t$ have norm $<1$.  
The assumptions 
on $\t$ and $\Psi$ imply easily that the integral of the kernel of 
either 
operator against either variable is bounded away {from} $1$, so the
operators have norm $<1$ in any $L^p$, $1\le p\le\infty$.
The asymptotics follow easily {from} the assumptions about $\t$
and $\Psi$, together with the trivial estimates
$$
0\le\r\le\t;\quad 0\le\s\le\Psi.\quad\qed
$$
\enddemo
One must make stronger assumptions to ensure finiteness of the 
integrals (8.6) and (8.7) and the relative energy density
$$
E=-\frac{c_1}{2\pi}\intr(\r-\r_0)\eta\,d\eta+
\frac{c_2}{2\pi}\intr(\s-\s_0)\z\,d\z.\tag 8.10
$$

\proclaim{Theorem 8.2} The integrals (8.6), (8.7) are
finite if $(\t-\t_0)$ and $(\Psi-\Psi_0)$ are integrable.

If $\eta(\t_0-\t)$ and $\z(\Psi_0-\Psi)$ are integrable, then $E$ is
finite if and only if (8.7) holds. If so, then $E\ge0$.
\endproclaim

\demo{Proof} Note that
$$
\r-\r_0=(\t-\t_0)(1-K\s_0)-\t K(\s-\s_0) ,\tag 8.11
$$ 
with a similar equation for $\s-\s_0$.  Analyzing this system as
above, we see that integrability of $\r-\r_0$ and $\s-\s_0$ 
follows from integrability of the data in the analogue of (8.9),
which follows in turn from integrability of $\t-\t_0$ and 
$\Psi-\Psi_0$.

Note that integrability of $\eta(\r_0-\r)$
at $-\infty$ follows immediately {from} (8.11) and
integrability of $\eta(\t_0-\t)$.
At $+\infty$ we observe {from} (8.11) that the necessary and sufficient 
condition is the integrability of $\eta K(\s-\s_0)$.  Assume that
$\z^2(\Psi-\Psi_0)$ is integrable at $-\infty$.  Then 
$\z^2(\s-\s_0)$ is also integrable at $-\infty$ and Lebesgue's Dominated
Convergence Theorem implies that
$$
\lim_{\eta\to+\infty}K(\s-\s_0)=\intr(\s-\s_0)\,d\z.\tag 8.13
$$
This proves the necessity of the conditions (8.7).

Conversely one can deduce the integrability of $\eta K(\s-\s_0)$ and
$\z K(\r-\r_0)$ {from} conditions (8.7) together with 
integrability of $\eta(\r_0-\r)$ and $\z(\s_0-\s)$.
To see this, note that the mean-value zero condition
allows us to replace the kernel of $K$ with the kernel
$$
\tilde k(\eta,\z)=\frac1\pi\bigg[\frac1{1+(\eta+\z)^2}-\frac1{1+\eta^2}.
\bigg]\tag 8.14
$$
The corresponding operator takes
$L^1(\R,(1+|z|)\,d\z)$ to $L^1(\R,(1+|\eta|)\,d\eta)$, since we can 
multiply $\tilde k$
by $(1+|\eta|)/(1+|z|)$, take the absolute value, and estimate the integral
with respect to $\eta$. 

Finally, to show positivity of the relative energy density we let
$g_1$ and $g_2$ denote the integrals in (8.6), so that
$$
\frac{dg_1}{d\eta}=\r,\quad \frac{dg_2}{d\z}=\s;\quad g_j\ge 0,
\quad g_j(\pm\infty)=0.\tag 8.15
$$
Integrating by parts in (8.10) we find that
$$
E=\frac{c_1}{2\pi}\intr g_1(\eta)\,d\eta-\frac{c_2}{2\pi}
\intr g_2(\z)\,d\z\ge 0.
\quad\qed 
$$
\enddemo

\ni
{\bf 9. THERMODYNAMICS}
\ss
We consider now the thermodynamics of the finite density perturbations
of the ground state that were constructed in the preceding section. 
We follow, with appropriate modifications, Yang and Yang$^5$ and 
Okhuma$^7$.  
Consider first the case of finite $L$, a state with a very large number 
of particles, and data $I,J$ as in (8.1).  Define
$$\align
&h(s)=s-\frac1\pi\int^\D_0[\tnin(s+\zeta(\xi'))
- \tnin(\eta(0)+\zeta(\xi'))]\,d\xi',\quad s\ge\eta(0);\tag 9.1\\
&k(t)=t-\frac1\pi\int^\D_0[\tnin(\eta(\xi')+t)
- \tnin(\eta(\xi')+\zeta(0))]\,d\xi',\quad t\ge\z(0).\endalign
$$
These are strictly increasing functions with the properties that
at the (discrete) values taken by the data the values are
$$
h(\eta(\xi))=I(\xi),\quad k(\z(\xi))=J(\xi).\tag 9.2
$$
The number $m$ of $\eta$ particles {from} our state with momenta in a
small interval $(\eta,\eta+d\eta)$ is proportional to the length of 
the corresponding $\xi$ interval, i.e.
if $\eta=\eta(\xi)$, $\eta+d\eta=\eta(\xi+d\xi)$ then
$$
m\approx\frac{L}{2\pi}d\xi. \tag 9.3
$$
The function $I$ has jumps that are positive integer multiples of
$L/2\pi$, so
$$
m\le n=\frac{L}{2\pi}(I(\xi+d\xi)-I(\xi)).
$$
If data $I$ at the $m$ jump sites in the interval $(\xi,
\xi+d\xi)$ is changed in such a way that $I$ remains strictly increasing and
the values of $LI/2\pi$ are integers, the effect on the total energy
of the corresponding state will be negligible: $O((d\eta)^2)$.   
The possible sites are indexed by the $n$ integers between 
$LI(\xi)/2\pi$ and $L(I\xi)/2\pi$.  Thus the contribution to entropy
{from} the interval is $\log{n\choose m}$.  

In the infinite length and infinite density limits the density functions
for the numbers $n$ and $m$ above have densities with respect to the
momentum $\eta$ given by $\r=d\xi/d\eta$ and $\r_t=dh/d\eta$ respectively.
As before $\r$ is the  particle density.  We take $\r_h=\r_t-\r$
to be the {\it hole density\/} and $\r_t$ to be the (formal) total density.
Similar considerations apply to $\z$ particles and give particle
density $\s$ and hole density $\s_h=\s_t-\s$.  In the notation of \S 8
the relations among these functions are
$$
\r=\t\r_t,\quad\r_h=(1-\t)\r_t;\qquad \s=\Psi\s_t,\quad\s_h=(1-\Psi)\s_t;
\tag 9.2
$$
$$
\r_t=1-K\s=1-K\Psi\s_t; \qquad\s_t=1-K\r=1-K\t\r_t.
$$

Returning to the interval $(\xi,\xi+d\xi)$, let $(\eta,\eta+d\eta)$
denote its image under $h\inv$.  
According to the preceding argument and Stirling's formula, the 
entropy density over $(\eta+d\eta)$ is approximately
$$\align
&\frac{L}{2\pi}[\r_t\log\r_t-\r\log\r-\r_h\log\r_h]\tag 9.3\\
&\quad =\frac{L}{2\pi}[\r\log(\r_t/\r)+\r_h\log(\r_t/\r_h)]\\
&\quad =-\frac{L}{2\pi}\r_t[\t\log\t+(1-\t)\log(1-\t)].\endalign
$$
A similar formula holds for particles of the second type.
Thus in the infinite length, infinite density limit the entropy 
per unit length is 
$$\align
S=&\frac1{2\pi}\intr[\r_t\log\r_t-\r\log\r-\r_h\log\r_h]
\,d\eta \tag 9.4\\
&+\frac1{2\pi}\intr[\s_t\log\s_t-\s\log\s-\s_h\log\s_h]\,d\z.
\endalign
$$
The entropy density for the ground state is zero.

At a given finite temperature $T$ the free energy density of a 
perturbation $(\r,\s)$ of the ground state relative to that of
the ground state is
$$
F_T(\r,\s)=E-T\,S\tag 9.5
$$
where the relative energy density $E$ is given by (8.9).  In order
for $E$ to be finite, we showed that the
relative particle densities must vanish:
$$
\intr(\r-\r_0)\,d\eta=0=\intr(\s-\s_0)\,d\zeta.\tag 9.6
$$
Thus the equilibrium solution should minimize $F$ subject to the
constraints (9.6).  

We consider briefly the formal argument for an equilibrium.  Equations
(9.2) relate variations in $\r_t$, $\r_h$, $\s_t$, and $\s_h$ to
variations in the particle densities $\r$, $\t$:
$$\align
\dot\r_t&=-K\dot\s,\quad\dot\r_h=-\dot\r-K\dot\s;\tag 9.7\\
\dot\s_t&=-K\dot\r,\quad\dot\s_h=-\dot\s-K\dot\r.\endalign
$$
The constraints (9.6) require
$$
\intr\dot\r\,d\eta=0=\intr\dot\s\,d\z.\tag 9.8
$$
Taking into account (9.7) and the symmetry of the operator $K$,
we can express the variation in the free energy as
$$\align
\dot F&= \intr\dot\r\,\big[-c_1\eta+T\log(\r/\r_h)+TK\log(\s_t/\s_h)\big]
\,d\eta\tag 9.9\\
&\quad+\intr\dot\s\,\big[c_2\z+T\log(\s/\s_h)+TK\log(\r_t/\r_h)\big]\,d\z.
\endalign
$$
Following Yang and Yang$^5$, one can introduce functions $\e_j$:
$$
e^{-\e_1/T}=\r/\r_h,\quad e^{-\e_2/T}=\s/\s_h.\tag 9.10
$$
and derive Euler-Lagrange equations in the form
$$\align
\e_1&=A_1 T-c_1\eta+TK\log(1+e^{-e_2/T}),\tag 9.11\\
\e_2&=A_2 T+c_2\z+TK\log(1+e^{-e_1/T}),\endalign
$$
where the $A_j$ are the Langrange multipliers that correspond to
(9.8).  

As noted in the introduction, 
there are two difficulties with this argument.  First,
equations (9.11) have no solution: taking derivatives one sees
that the $\e_j$ should be decreasing functions, so the functions
$\log(1+e^{-e_j/T})$ are increasing and bounded above, but 
they are taken by $K$ to decreasing functions and thus
one needs $\e_1\le-c_1\eta+O(1)$ at $+\infty$.
Therefore $\log(1+e^{-e_1/T})$ is nonnegative and grows like $\eta$
at $+\infty$, so $K\log(1+e^{-e_1/T})$ cannot be defined.
Second, it is not clear that the free energy is
bounded below.

The first difficulty is easily dealt with.  We may assume that $\dot r$
and $\dot s$ are test functions, and $(9.9)$ simply means that they
are derivatives of test functions.  Therefore the condition $\dot F=0$
simply means that the derivatives of the two bracketed terms vanish.
Now $K$ anti-commutes with derivation, so the correct
form of (9.11) is
$$
\frac{d\e_1}{d\eta}=-c_1+K\bigg(\Psi\frac{d\e_2}{d\z}\bigg);\quad
\frac{d\e_2}{d\z}=c_2+K\bigg(\t\frac{d\e_1}{d\eta}\bigg).
$$
Combining this with
$$
\t=\frac{e^{-\e_1/T}}{1+e^{-e_1/T}},\quad
\Psi=\frac{e^{-\e_2/T}}{1+e^{-e_2/T}}\tag 9.13
$$
yields a well-behaved system of equations for the thermal equilibrium
particle distribution $(\r,\s)=(\r^{(T)},\s^{(T)})$.

To deal with the second difficulty we prove the following.

\proclaim{Theorem 9.1} The free energy functional $F_T$ is 
convex and bounded below by $-CT^2$, where $C$ is a constant
that depends only on the velocities $c_j$.\endproclaim

\demo{Proof} {From} (9.9), the second derivative of the free energy for the
variation $(\dot\r,\dot\s)$ is
$$\align
T&\intr\dot\r\bigg[\frac{\dot\r}{\r}-\frac{\dot\r_h}{\r_h}+K\bigg(
\frac{\dot\s_t}{\s_t}-\frac{\dot\s_h}{\s_h}\bigg)\bigg]\,d\eta\tag 9.14\\
&\qquad+T\intr\dot\s\bigg[\frac{\dot\s}{\s}-\frac{\dot\s_h}{\s_h}+K\bigg(
\frac{\dot\r_t}{\r_t}-\frac{\dot\r_h}{\r_h}\bigg)\bigg]\,d\zeta\\
&=T\intr\bigg[\dot\r\bigg(\frac{\dot\r}{\r}-\frac{\dot\r_h}{\r_h}\bigg)
+(K\dot\s)\bigg(\frac{\dot\r_t}{\r_t}-\frac{\dot\r_h}{\r_h}\bigg)
\bigg]\,d\eta\\
&\qquad+T\intr\bigg[\dot\s\bigg(\frac{\dot\s}{\s}-\frac{\dot\s_h}{\s_h}\bigg)
+(K\dot\r)\bigg(\frac{\dot\s_t}{\s_t}-\frac{\dot\s_h}{\s_h}\bigg)
\bigg]\,dz.\endalign
$$
We use (9.7) and reorganize to find that the second derivative is
$$
T\intr\bigg(\frac{\sqrt{\r_t}}{\sqrt{\r\r_h}}\,\dot\r +\frac{\sqrt{\r}}
{\sqrt{\r_t\r_h}}\,K\dot\s\bigg)^2\, d\eta
+T\intr\bigg(\frac{\sqrt{\s_t}}{\sqrt{\s\s_h}}\,\dot\s +\frac{\sqrt{\s}}
{\sqrt{\s_t\s_h}}\,K\dot\r\bigg)^2\,d\zeta.\tag 9.15
$$

To prove boundedness we introduce the function
$$
G(s)=G(1-s)=-s\log s-(1-s)\log(1-s),\quad 0\le s\le 1.
$$
Then $G(\t)=G(|\t_0-\t|)$ and the entropy 
density can be written
$$
2\pi S(\r,\s)=\intr\r_t G(\t)\,d\eta+\intr\s_t G(\Psi)\,d\z.\tag 9.16
$$
For the first integral we partition the line into three sets
$$\align
I_1&=\{\eta: G(\t)\le |\eta(\t_0-\t)|/\delta\},\tag 9.17\\
I_2&=\{\eta: |\t_0-\t|\ge 1/2,\ G(\t)>|\eta(\t_0-\t)|/\delta\},\\
I_3&=\{\eta: |\t_0-\t|< 1/2,\ G(\t)>|\eta(\t_0-\t)|/\delta\},
\endalign
$$
where the constant $\delta$ is to be chosen.
Note that $0\le G(t)\le\log 2$ so $\eta\in I_2$ implies 
$|\eta|<2\delta\log2$.  Also
$$
-\log t\ge -2t\log t\ge G(t),\quad 0\le t\le1/2
$$
so $\eta\in I_3$ implies $|\t_0-t|<\exp(-|\eta|/\delta)$.  Now
$-t\log t$ is increasing on the interval $[0,1/e]$, so we conclude
that $\eta\in I_3$ implies that either $|\eta|/\delta <1$ or
$$
G(\t)=G(|\t_0-\t|)\le- 2|\t_0-\t|\log|\t_0-\t|\le 2\frac{|\eta|}{\delta}
e^{-|\eta|/\delta}.
$$
Since $1<2\log2$ we can combine these estimates and obtain
$$\align
\intr\r_tG(\t)\,d\eta&\le\frac1\delta\intr\r_t(\t_0-\t)\eta\,d\eta+4\delta\log2
+2\intr\frac{|\eta|}\delta e^{-|\eta|/\delta}\,d\eta\tag 9.18\\
&=\intr\r_t(\t_0-\t)\eta\,d\eta+4\delta\log2+4\delta.\endalign 
$$
We claim that
$$
\intr\r_t(\t_0-\t)\eta\,d\eta\le\intr(\r_0-\r)\eta\,d\eta
+\intr (\s_0-\s)\z\,d\z.\tag 9.19
$$
In fact (8.11) can be rewritten
$$
\r_t(\t_0-\t)=\r-\r_0+\t_0K(\r-\r_0).\tag 9.20
$$
Using $\langle\ ,\ \rangle$ to denote the standard inner product,
we want to prove that 
$$\align
\intr\t_0K(\s-\s_0)\eta\,d\eta&=\la K(\s-\s_0)\,,\,\t_0\eta\ra\tag 9.21\\
&\le\la\s_0-\s\,,\,\z\ra=-\la Dg_2\,,\,z\ra =\la g_2\,,\,1\ra,\endalign
$$
where $D$ denotes differentiation and $g_2$ is the function in
(8.15).  We use the fact that derivation anti-commutes with $K$ 
and integrate by parts to obtain
$$\align
\la K(\s-\s_0)\,,\,\t_0\eta\ra&=\la KDg_2\,,\,\t_0\eta\ra=
-\la DKg_2\,,\,\t_0\eta\ra\\
&=\la Kg_2\,,\,D(\t_0\eta)\ra
=\la g_2\,,\,K\t_0\ra.\endalign
$$
Now $0\le K\theta_0\le 1$ and $g_2\ge 0$, so this 
proves (9.19).  Using (9.18), (9.19), and the analogous
results for $\s$, we obtain 
$$\align
F_T(\r,\s)&\le\bigg(c_1-\frac{2T}{\delta}\bigg)\intr(\r_0-\r)\eta\,d\eta
+4\delta(1+\log2)\tag 9.22\\
&+\bigg(-c_2-\frac{2T}{\delta}\bigg)\intr(\s_0-\s)\zeta\,d\zeta
+4\delta(1+\log2).\endalign
$$
We take $\delta=2T/(c_1-c_2)$ and obtain
$C=8(1+\log2)/(c_1-c_2)$.\quad\qed\enddemo
\ms
\ni{\bf ACKNOWLEDGMENTS}
\ss
This research was supported by National Science
Foundation grants DMS-9213595 and DMS-9501233.
\ms
\ss\ni
$^1$  M.~Wadati, and K.~Okhuma, `` Bethe States for the Quantum 
Three Wave Interaction Equation,'' J.~ Phys.~Soc.~Japan
{\bf 53}, 1229-1237 (1984).

\ss\ni
$^2$  P.~P.~Kulish and N.~Reshetikhin, ``Diagonalization of GL(N) 
invariant transfer matrices and quantum N-wave system (Lee Model),''
J.~ Phys.~A: Math.~Gen.~{\bf 16}, L591-L596 (1983).

\ss\ni
$^3$  V.~E.~Korepin, ``Direct calculation of the S Matrix in the massive 
Thirring Model,'' Theor.~and Math.~Phys.~ {\bf 41},
168-189 (1979).

\ss\ni
$^4$   H.~Bergknoff and H.~B.~Thacker, ``Structure and solution of
the massive Thirring model,'' Phys.~Rev.~D {\bf 19}, 3666-3681 (1979).

\ss\ni
$^5$  C.~N.~Yang and C.~P.~Yang, ``Thermodynamics of a one-dimensional
system of bosons with repulsive delta-function interaction,''
J.~Math.~Phys.~{\bf 10} (1996), 1115-1122. 

\ss\ni
$^6$  K.~Ohkuma and M.~Wadati, 
``Quantum Three Wave Interaction Models,''
J.~ Phys. Soc.~Japan {\bf 53}, 2899-2907 (1984).

\ss\ni
$^7$  K.~Ohkuma, ``Thermodynamics of the Quantum Three Wave Interaction 
Model,''  J.~ Phys.~Soc.~Japan {\bf 54}, 2817-2828 (1985).

\ss\ni
$^8$  E.~H.~Lieb, and W.~Liniger, ``Exact analysis of an interacting
Bose gas, I. The general solution and the ground state,''
Phys.~Rev.~{\bf 130}, 1605-1615 (1963).

\ss\ni
$^9$  V.~E.~Korepin, N.~M.~Bogoliubov, and V.~E.~Izergin,  {\sl
Quantum Inverse Scattering Method and Correlation Functions}  
(Cambridge University Press, Cambridge, 1993).

\ss\ni
$^{10}$  H.~Bethe, ``Zur Theorie der Metalle. I. Eigenwerte und Eigenfunktionen
der linearen Atomkette,'' Z.~Phys.~{\bf 71}, 205-226 (1931).

\ss\ni
$^{11}$  E.~H.~Lieb, ``Exact analysis of an interacting Bose gas, II. 
The excitation spectrum,'' Phys.~Rev.~{\bf 130}, 1616-1624 (1963).

\ss\ni
$^{12}$  C.~N.~Yang and C.~P.~Yang, ``One-dimensional chain of anisotropic
spin-spin interactions I. Proof of Bethe's hypothesis for ground state
in finite system,'' Phys. Rev.~{\bf 150}, 321-327 (1966).

\ss\ni
$^{13}$  V.~E.~Zakharov and S.~V.~Manakov, ``Resonant interactions
of wave packets in nonlinear media,'' JETP {\bf 18}, 243-245 (1973).

\ss\ni
$^{14}$  R.~Beals, and D.~H.~Sattinger, ``On the Complete Integrability of
Completely Integrable Systems,'' Comm.~Math.~Phys.~{\bf 138},
409-436 (1991).

\ss\ni
$^{15}$  R.~Beals, D.~H.~Sattinger, and E.~Williams, ``Quantization of the
three wave interaction,'' Technical Report, www.math.umn.edu/$\sim$dhs.

\end

[K] Kulish, P.~P.: Quantum nonlinear wave interaction system.
{\sl Physica} {\bf 18D}, 360-364 (1986)

[NMPZ] Novikov, S., Manakov, S.~V., Pitaevskii, L.~P., and Zakharov,
V.~E.:{\sl Theory of Solitons\/}.
Consultants Bureau, New York, 1984

[PP] Peng, Y.~and Pu, F.: One-dimensional quantum three-wave interaction 
model with two fermionic fields and one bosonic field. {\sl J.~ Math.~Phys.},
{\bf 29}, 400-402 (1988)